\documentclass[times,twocolumn,final]{elsarticle}
\usepackage{nicefrac}
\usepackage{framed,multirow}
\usepackage{framed,multirow}
\usepackage{booktabs}
\usepackage[export]{adjustbox}
\usepackage{amsmath,amssymb,amsmath}
\usepackage{algorithm,algorithmic}
\usepackage{multirow}
\usepackage{siunitx}
\usepackage{blindtext}
\usepackage[utf8]{inputenc}
\usepackage{commath}
\usepackage{graphicx}
\usepackage{url}
\usepackage[twoside,
paperwidth=210mm,
paperheight=280mm,
textheight=635.35pt,
textwidth=467.75pt,
inner=32pt,
outer=43pt,
top=65pt,
bottom=40pt,
headheight=12pt,
headsep=12pt,
footskip=16pt,
footnotesep=28pt plus 2pt minus 6pt,
columnsep=18pt]{geometry}
\usepackage{tikz}
\newtheorem{theorem}{Theorem}[section]
\newtheorem{corollary}[theorem]{Corollary}
\begin{document}

\begin{abstract}
Inspired by the first-order method of Malitsky and Pock, we propose a new variational framework for compressed MR image reconstruction which introduces the application of a rotation-invariant discretization of total variation functional into MR imaging while exploiting BM3D frame as a sparsifying transform. {In the first step, we provide theoretical and numerical analysis establishing the exceptional rotation-invariance property of this total variation functional and observe its superiority over other well-known variational regularization terms in both upright and rotated imaging setups.} Thereupon, the proposed MRI reconstruction model is presented as a constrained optimization problem, however, we do not use conventional ADMM-type algorithms designed for constrained problems to obtain a solution, but rather we tailor the linesearch-equipped method of Malitsky and Pock to our model, which was originally proposed for unconstrained problems.  As attested by numerical experiments, this framework significantly outperforms various state-of-the-art algorithms from variational methods to adaptive and learning approaches and in particular, it eliminates the stagnating behavior of a previous work on BM3D-MRI which compromised the solution beyond a certain iteration.  
\end{abstract}

\begin{keyword}Magnetic Resonance Imaging (MRI), Compressed Sensing, Variational Image Processing, Iterative Image Reconstruction, Denoising, First-Order Methods.
\end{keyword}

\begin{frontmatter}
\title{Compressed MRI reconstruction exploiting a rotation-invariant \\total variation discretization}

\author[1]{Erfan~Ebrahim~Esfahani\corref{cor1}}
\ead{ebrahim.esfahani@ut.ac.ir}

\cortext[cor1]{Corresponding author.}

\author[1]{Alireza~Hosseini}
\ead{hosseini.alireza@ut.ac.ir}
\address[1]{School of Mathematics, Statistics and Computer Science, College of Science, University of Tehran, Tehran, P.O. Box 14115-175, Iran}
\end{frontmatter}

\section{Introduction}
Magnetic
resonance imaging is a non-ionizing and non-invasive medical imaging modality that provides outstanding visual representation of biological tissues and anatomical functions. However, it has its own drawbacks, most notably the slow acquisition process, which weighs on both the patient (a long time spent inside an enclosed magnet) and the clinic (extended power consumption, wear and tear, etc). Therefore, accelerated imaging has been  a subject of interest among the MRI research community over the past decade.
 The underlying theory of compressed sensing
\cite{CS}
offers a solution to this problem: instead of sampling the full k-space, a limited number of samples are taken and then a nonlinear reconstruction method exploiting optimization techniques is employed
\cite{LDP,Ma MRI,FCSA,CS MRI,TGV MRI,TGVSH}. The common thread in these classical approaches is the combination of a variational penalty term with a fixed sparsifying transform which models prior knowledge of the solution. 


Recent trends in compressed MRI have shifted focus to adaptive frameworks where the transform is image-specific. This approach allows for the sparsifying transform to be tailored to each specific image in a patchwise sense and hence leads to much better sparse representations than fixed transforms such as wavelets, contourlets etc (collectively called X-let transforms).
 This basic idea, in various formulations and methodologies, has been explored, for example in \cite{DLMRI,TLMRI,FRIST,BM3D-MRI,FDLCP,GBRWT}. What characterizes all these methods is the fact that the adaptation involves a \textit{single} image.

 Another paradigm gaining popularity in data science and specifically in MRI compressed sensing is the so-called deep learning. One of the characteristics of this approach that differentiates it from the methods that we have discussed so far, is the fact that deep architectures learn all the necessary components from a large set of training examples, involving \textit{hundreds} or potentially \textit{thousands} of examples. However, one could argue that so far this emerging paradigm, although remarkably fast in online reconstructions, has not shown outstanding improvements over decent single-image-based approaches \cite{DAGAN}. For instance, the deep architecture in \cite{Deep ADMM} was observed to be slightly outperformed by the single-image-based method of \cite{BM3D-MRI} in \cite{Deep ADMM} and in \cite{VN} the deep architecture was only compared with baseline classical methods. Another notable drawback of deep learning is the tedious offline training procedures which could take hours \cite{Morteza}, or even days on current hardware systems.

In this paper, {as the first step, we establish the isotropy of a recent discretization of the total variation functional both theoretically and numerically.} Then, we propose a novel variational framework that introduces this rotation-invariant total variation functional into compressed MRI and at the same time exploits transform-domain sparsity by means of BM3D frames. The proposed MR image reconstruction model is expressed as a constrained minimization problem, however, we choose not to solve this model using popular ADMM-based methods designed for constrained problems but rather, we adapt a primal-dual algorithm with linesearch to solve our model which was originally proposed for unconstrained problems. This framework is shown to outperform various state-of-the-art methods of the literature and in particular, eliminates the stagnating behavior of a previous algorithm on BM3D-MRI \cite{BM3D-MRI} which degraded the reconstructed image beyond a certain, a priori unknown iteration.  

The rest of the paper is organized as follows. Section \ref{sec background} briefly reviews the related background and presents the proposed framework. Numerical experiments are conducted in Section \ref{sec exp} and final remarks are made in Section \ref{sec concl}.

\section{Material and methods } \label{sec background}
\subsection{A primal-dual algorithm with linesearch}
Suppose that 
$X$ is a finite-dimensional real vector space equipped with an inner product 
$\langle \cdot,\cdot\rangle_X$
and Euclidean norm
$||\cdot||_2=\sqrt{\langle \cdot,\cdot\rangle_X}.$ For any convex subset 
$C\subset X$ {
 the\textit{ indicator} function of $C$ is defined by $\delta_C(x) := 0$ if $x\in C$ and $\delta_C(x):=\infty$ otherwise. The function $\delta$ which can attain $+\infty$ as a value, is an example of extended real-valued functions  \cite[Chapter 2]{Beck}.
}
 Assume that
$f:X\rightarrow[-\infty,+\infty]$
is an extended real-valued function. The \textit{convex conjugate} for $f$ at
$\bar{x}\in X$ is defined by$
f^*(\bar{x}) := \max_x \,\langle x,\bar{x}\rangle - f(x). $



Now, suppose that $Y$ is another real vector space with inner product $\langle \cdot,\cdot\rangle_Y$ and the  induced norm $||\cdot||_2 = \sqrt{\langle \cdot,\cdot \rangle_Y}$.
Let $K:X\rightarrow Y$ be a bounded linear operator. 
The adjoint 
 of $K$ is defined (with a small but common abuse of notation) as an operator $K^*$ for which the equality 
$\langle Kx,y \rangle_Y = 
\langle x, K^*y \rangle_X $ holds for all $x$ and $y$.  
First-order methods \cite{Beck,Cham-Pock,MP,ASGARD,GADMM} usually focus on the problem
\begin{equation}\label{min max prob}
\min_x \max_y \,\langle Kx,y \rangle_Y + g(x) - f^*(y),
\end{equation}
where extended functions $f:X\rightarrow [0,\infty]$ and 
$g:Y\rightarrow [0,\infty]$
are proper, closed and convex. The method we employ to solve
(\ref{min max prob}) is the primal-dual algorithm with linesearch proposed by Malitsky and Pock \cite{MP}. Briefly, this algorithm is based on the Chambolle-Pock algorithm \cite{Cham-Pock} and equips that method with a linesearch in the dual variable update step. Inclusion of linesearch remarkably improves the performance of this method while only slightly increasing the workload. 
Assuming appropriate positive constants $\delta$ and $\beta$ are chosen, the iteration steps of \cite{MP} for the template problem (\ref{min max prob}) are given by
\begin{align}
& x^k = \text{prox}_{\tau_{k-1}g}(x^{k-1}-\tau_{k-1}K^*y^{k-1}), \label{x} \\
&\text{Choose } \tau_k \in [\tau_{k-1}, \tau_{k-1}\sqrt{1+\theta_{k-1}}], \notag \\
&\theta_k = \frac{\tau_k}{\tau_{k-1}}, \label{teta}  \\
    &\bar{x}^k = x^k + \theta_k(x^k-x^{k-1}),\notag\\
    & y^k = \text{prox}_{\beta \tau_k f^*} (y^{k-1} + \beta \tau_k K \bar{x}^k), \label{y}\\
   &\text{if } \sqrt{\beta}\tau_k||K^*(y^{k}-y^{k-1})||_2 \leq \delta||y^k-y^{k-1}||_2 \notag \\
   &\text{return to }(\ref{x}),  
   \notag\\\ &\text{otherwise set }
   \tau_k = \mu \tau_k \text{ and return to } (\ref{teta}).
   \notag
\end{align}
 Henceforth we shall refer to this method as \textit{the Malitsky-Pock algorithm}. Definition and analysis of the \textit{proximal} (or \textit{proximity}) mapping 
 of a function $f$ denoted by $\text{prox}_f$ can be found in \cite[Chapter 6]{Beck}. The convergence of the algorithm to a solution for (\ref{min max prob}) is proved in \cite{MP} under standard conditions.

\renewcommand{\algorithmicrequire}{\textbf{Initialization:}}
  \renewcommand{\algorithmicensure}{\textbf{Output:}}

\subsection{Variational regularization}
Let $z \in Z_m = (\mathbb{R}^{n \times n})^m$ be an m-tuple of images each one of size $n \times n$. For instance, for $m=1$, $z$ is simply a size-$n\times n$ image and for $m=2$, $z=(z_1,z_2)$ is a pair of such images. For $1\leq p <\infty$ the space $Z_m$ is equipped with $l_{p,2}$-norm as 
\begin{equation}\label{lp norm}
||z||_{p,2} :=\bigg( \sum_{i,j=1}^n \big( \sqrt{\sum_{d=1}^m|z_d(i,j)|^2} \big)^p \bigg)^{\frac{1}{p}} ,\\
\end{equation}
where we agree to use $||\cdot||_2$ in lieu of $||\cdot||_{2,2}$ and for $p=\infty$ 
\begin{equation}\label{linf norm}
||z||_{\infty , 2} := \max_{i,j} \sqrt{\sum_{d=1}^{m}|z_d(i,j)|^2}.
\end{equation}
For convenience we set $Z_1 = U$ and $Z_2=V$.

Let $u\in U$. The \textit{total variation} of $u$ is usually {approximated} by
\begin{equation}\label{TVi}
\text{TV}(u) = ||Du||_{1,2} =\sum_{i,j=1}^n ||D u (i,j)||_{2},
\end{equation}
 where $Du = (D_1u,D_2u)$ with 
 $D_1 u(i,j) = u(i+1,j)-u(i,j)$ and $D_2 u(i,j)=u(i,j+1)-u(i,j)$ where Neumann boundary conditions are assumed \cite{Condat}.
Equation (\ref{TVi}) is sometimes referred to as the \textit{isotropic} TV since it is a straightforward discretization of 
\begin{equation}\label{TV cont}
\int_{\mathbb{R}^2} |\nabla f(x,y)|\,dxdy,
\end{equation}
which is known to be rotation-invariant for all functions $f$ with some regularity conditions \cite{Condat}. However, the discretization (\ref{TVi}) is far from being isotopic in discrete domain (see Fig.\,\ref{fig1}). A generalization of the total variation functional called \textit{total generalized variation} (TGV) was introduced in \cite{TGV}, which was given by 
$$ \text{TGV}(u) = \min_{v\in V} \alpha_1 ||Du -v||_{1,2} + \alpha_0||\mathcal{E}v||_{1,2}, $$
where $\mathcal{E}:V \rightarrow Z_4$ is a symmetrized derivative \cite{TGV}.
 This modification somewhat improves the performance of TV, but unfortunately it is not isotropic either (see again Fig. \ref{fig1}). 

\subsection{A rotation-invariant total variation} \label{RITV intro}

Resorting to convex analysis, it is not difficult to see that the \textit{primal} formulation (\ref{TVi}) has the equivalent \textit{dual} formulation 
\begin{equation}\label{TVi dual}
\text{TV}(u) = \max_{v\in V} \, \big\{\langle Du,v \rangle:\, ||v||_{\infty,2} \leq 1 \big\}.
\end{equation}  
The pixel intensity value is assumed to be located at the center of the pixel whereas{ a finite difference value such as $u(i+1,j) - u(i,j)$ is considered to be an approximation of the partial derivative at $(i+\frac{1}{2},j)$, that is, at the edge between the two horizontally adjacent pixels.} Similarly, $u(i,j+1) - u(i,j)$ is located at $(i,j+\frac{1}{2})$. A potential explanation for lack of rotation-invariance in TV is the fact that dual image intensities $v_1(i,j)$ and $v_2(i,j)$ corresponding to $D_1u(i,j)$ and $D_2u(i,j)$, present in the constraint $||u||_{\infty,2}\leq 1$, are also located at different positions $(i+\frac{1}{2},j)$ and $(i,j+\frac{1}{2})$ whereas they would have been expected at pixel center $(i,j)$. To correct this half-pixel shift, in \cite{Condat} the dual intensities $v_1(i,j)$ and $v_2(i,j)$ 
 are constrained to satisfy $||v||_{\infty,2}\leq 1$ not only on the edges (which is already done in (\ref{TVi dual})) but also at the pixel centers. More precisely, a new discretization for (\ref{TV cont}), which we henceforth refer to as \textit{rotation-invariant total variation} (RITV), is formulated:
\begin{equation} \label{RITV dual}
    \begin{split}
    \text{RITV} (u)&= \max_{v \in V}\, \big\{ \langle  Du,v \rangle: 
    ||L_{\updownarrow}v ||_{\infty,2} \leq 1, \, 
    \\ &|| L_{\leftrightarrow}v||_{\infty,2} \leq 1, \,
    || L_{\bullet} v||_{\infty,2} \leq 1,\, 
    ||L_+v||_{\infty,2} \leq 1
     \big\},
    \end{split}
    \end{equation}
    where
    \begin{equation}\label{L operators}
        \begin{split}
         L_{\updownarrow} v_1 (i,j) &= v_1(i,j),\\
        L_{\updownarrow}v_2 (i,j) & = \frac{1}{4}
        \big( v_2(i,j) + v_2(i,j-1) + v_2 (i+1,j) \\&+ v_2(i+1,j-1)   \big),\\
        L_{\leftrightarrow} v_1 (i,j) & = \frac{1}{4} \big(  v_1(i,j) + v_1(i-1,j) + v_1 (i,j+1) \\& + v_1(i-1,j+1)  \big),\\
         L_{\leftrightarrow} v_2(i,j) & = v_2(i,j), \\
         L_{\bullet} v_1 (i,j) & = \frac{1}{2}
        \big( v_1(i,j) + v_1(i-1,j) \big),\\
         L_{\bullet} v_2 (i,j) & = \frac{1}{2}\big( v_2(i,j) + v_2(i,j-1) \big),\\
         L_+ v_1 (i,j) &= \frac{1}{2} \big (v_1(i,j)+v_1(i,j+1  \big) ,\\
         L_+ v_2(i,j) &= \frac{1}{2}
         \big( v_2(i,j) + v_2(i+1,j) \big).
        \end{split}
        \end{equation}
       
    {    We assume zero boundary conditions, that is,
  for any $(i,j)\in \{1,2,\cdots,n\}\times\{1,2,\cdots,n\}$, $ v_1(0,j)=v_2(i,0)=v_1(n,j)=v_2(i,n)=0$. We also set 
              $L_{\updownarrow} v_1(n,j)= L_{\updownarrow}v_2(n,j)=0,$ $L_{\leftrightarrow} v_1(i,n)= L_{\leftrightarrow}v_2(i,n)=0,$ $
              L_{+}v_1(i,n)=L_{+}v_2(i,n)=0$ and $
              L_{+}v_1(n,0)=L_{+}v_2(n,0)=0.
              $}

    As with (\ref{TVi}), a primal formulation for the dual form of RITV (\ref{RITV dual}) can be achieved through convex analysis \cite{Condat}:
    \begin{equation} \label{RITV primal}
    \begin{split}
    \text{RITV}(u) &= \min_{v_{\updownarrow},v_{\leftrightarrow},v_{\bullet},v_+\in V} \sum_{s\in \{\updownarrow,\leftrightarrow,\bullet,+ \} } ||v_s||_{1,2}\\
    &\text{s.t. } \sum_{s\in \{\updownarrow,\leftrightarrow,\bullet,+ \}} L^*_sv_s - Du = 0.
       \end{split}
    \end{equation}
      \begin{figure}
\centering\includegraphics[width=0.9\linewidth]{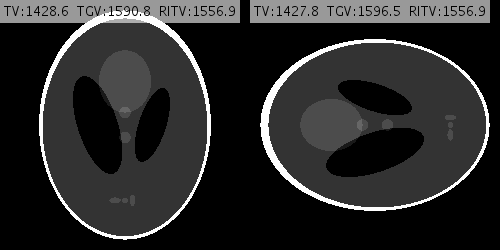}
\caption{A Shepp-Logan phantom of size $250\times 250$, upright (left) and rotated by $90^{\circ}$  counterclockwise (right). Inserted to top of each image are the values for TV, TGV and RITV. RITV retains the same value (up to machine precision) after rotation while TV and TGV fail to do so. The value of TV was obtained from (\ref{TVi}) while TGV was computed by 200 iterations of the Chambolle-Pock algorithm \cite{Cham-Pock} and RITV was calculated through 200 iterations of Algorithm 2 in \cite{Condat}. }
             \label{fig1}
             \end{figure}
   The isotropy of RITV, exemplified in Fig. \ref{fig1}, will be established in Subsection \ref{theory}.
\subsection{Proposed framework}\label{sec proposed}
    \subsubsection{The model}
  

Based on the observations above, we propose the following minimization problem for compressed MR image reconstruction (we declare the index set $S:=\{\updownarrow,\, \leftrightarrow,\,\bullet,\,+ \} $ for simplicity):

\begin{equation} \label{proposed model}
\begin{split}
\min_{u,\{v_s:s\in S\}}
\frac{1}{2}||&\mathcal{F}_\mathcal{M}u-b||_2^2 + \eta||\Phi u||_0 + 
\lambda\sum_{s\in S} ||v_s||_{1,2}\\
    &\text{s.t. } \sum_{s\in S} L^*_sv_s - Du = 0,
\end{split}
\end{equation}
where $u\in U$ is the sought-after MR image, $\Phi$ is the \textit{block matching 3D} (BM3D) analysis frame \cite{BM3D 1, BM3D 2}, $\mathcal{F}_{\mathcal{M}}$ is the undersampled Fourier operator 
defined as $\mathcal{F}_{\mathcal{M}} u := \mathcal{M} \odot \mathcal{F}(u)$ where $\mathcal{F}$ is the 2-D fast Fourier transform, $\mathcal{M}$ is the k-space sampling mask with ones at sampled frequencies and zeros at unsampled locations and $\odot$ is componentwise multiplication. We further assume that the sampling rate $\nicefrac{\# \text{samples} }{n^2}$ is considerably small (highly undersampled k-space) and $b\in \mathbb{C}^{n\times n}$ is the partially scanned k-space.
The constants $\eta,\,\lambda \in \mathbb{R}$ are regularization parameters. The zero-filling solution is defined by $u_\text{zf} :=\mathcal{F}^{-1}(b)$.  

Problem (\ref{proposed model}) falls into the class of constrained optimization problems which are usually solved by ADMM-type algorithms (that is, ADMM or its variants). However, we would like to exploit the Malitsky-Pock algorithm which primarily deals with unconstrained problems. From the result in \cite[Example 4.2]{Beck}, it is seen that  $\delta^*_{\{0\}}(h) = \max_{\sum_s L_s^*v_s-Du=0} \langle \sum_s L^*_s v_s -Du,h\rangle = 0$ for all $h \in V$. On the other hand, applying the definition of convex conjugate to $\delta^*_{\{0\} }(h)$ and noting that $\delta_{\{0\}} = \delta^{**}_{\{0\}}$ (since $\delta_{\{0\}}$ is proper, closed and convex), we obtain 
$\delta_{\{0\}}(\sum_s L_s^*v_s-Du) = \max_h \langle \sum_s L_s^*v_s-Du, h \rangle$. We can now write the following equivalent reformulations for
(\ref{proposed model}):

 \begin{equation}
 \begin{split}
 \min_{u,\{v_s:s\in S\}}
 \frac{1}{2}||&\mathcal{F}_\mathcal{M}u-b||_2^2 + \eta||\Phi u||_0 + 
 \lambda\sum_{s\in S} ||v_s||_{1,2} \\&+  \delta_{\{0\}}(\sum_{s\in S}L^*_sv_s - Du)\\ \iff
 \min_{u,\{v_s: s\in S\}}\, &\max_{h\in V, r\in\mathbb{C}^{n\times n}}\, \eta||\Phi u||_0 + \lambda \sum_{s\in S} ||v_s||_{1,2} \\+ \langle \mathcal{F}_{\mathcal{M}}u -b,& r\rangle - \frac{1}{2} ||r||_2^2 
 + \langle \sum_{s\in S} L^*_sv_s -Du, h \rangle.\label{proposed model 2}
 \end{split}
 \end{equation}
 Setting $x:=(u,v_{\updownarrow},v_{\leftrightarrow},v_{\bullet}, v_+)^\top$, $y:=(r,h)^\top$,
 \begin{align}
    g(x) & :=\eta||\Phi u||_0 + \lambda\sum_{s\in S} ||v_s||_{1,2},\label{g}\\
    f^*(y) &:= \frac{1}{2}||r||_2^2 +\langle b,r \rangle ,\label{f*}
     \end{align}    
and 
 \begin{equation}
    K  := 
    \begin{pmatrix}
    \mathcal{F}_{\mathcal{M}} & 0 & 0 & 0 & 0\\
    -D & L^*_{\updownarrow} & L^*_{\leftrightarrow} & L^*_{\bullet} & L^*_+ 
    \end{pmatrix},
 \end{equation} 
  the iterations of the Malitsky-Pock method for (\ref{proposed model 2}) may be obtained. However, before we develop the algorithm, we wish to make an observation about the isotropy of RITV.
 { \subsubsection{The theory} \label{theory}}
 
 An expected property of discrete models as approximations to continuous-domain models is preservation of the features present in the original continuous formulation. In this subsection we observe that RITV preserves the exceptional continuous-domain property of rotation-invariance, and in fact we observe more: under a norm-preserving condition in compressed MRI, the RITV reconstruction remains isotropic; that is, after rotating the zero-filled solution  $u_{\text{zf}}$ by $\theta=90^0, 180^0, \cdots$, the RITV-regularized result will not change and it will only be a $\theta$-rotated version of the same output obtained from the unrotated $u_{\text{zf}}$, as is expected from an isotropic model.  The following theorem states this:

  \begin{theorem} \label{theorem}
  	For any two dimensional image $A,$ let $\mathcal{R}A$ be the rotated version of the image by $90^0$. Consider the proposed model (\ref{proposed model}) and set $\eta=0$. Assume that the sampling mask has the property that for the image $A,\, \|\mathcal{M}\odot(\mathcal{R}A)\|_2=\|\mathcal{M}\odot A\|_2$ and $b\in\mathbb{C}^{n\times n}$ is the partially scanned k-space which is determined by $\mathcal{M}$ with zero-filled solution $u_{\text{zf}}$. If $u$ is a solution of  (\ref{proposed model}) with scanned k-space $b$ that gives the zero-filled solution $u_\text{zf}$, then $\mathcal{R}u$ is a solution of (\ref{proposed model}) with a scanned k-space that gives the zero-filled solution $\mathcal{R} u_{\text{zf}}.$ 
  \end{theorem}

The theorem is proved in \ref{proof of theorem}.
In the proof, the rotation-invariance property discussed above is also established (which was observed numerically in Fig. \ref{fig1}): 
\begin{corollary}
 Let $u$ be any two-dimensional gray-scaled digital image (which does not necessarily need to be an MR image) and assume that $\mathcal{R}u$ is the same image rotated by $90^\circ$. Then $\text{RITV}(\mathcal{R}u)=\text{RITV}(u).$
\end{corollary}

 We remark that the purpose of the norm-preserving assumption $\|\mathcal{M}\odot(\mathcal{R}A)\|_2=\|\mathcal{M}\odot A\|_2$ in Theorem \ref{theorem} is purely theoretical. In practice, designing a k-space trajectory that preserves this property might be tedious and of little interest in clinical application. In Subsection \ref{TV, TGV and RITV isotropy} we shall observe that although, for example, the clinically popular Cartesian sampling may violate this assumption, the RITV solution is still remarkably isotropic, and definitely more so than TV and TGV solutions.

 \subsubsection{The algorithm}
 {The $x$-subproblem}:  
 Since $g$ is separable the subproblem 
 (\ref{x})
  decouples into a $u$ component and four $v_s$ components, one for each $s\in S $. 

 The update for $u$ is given by
 \begin{align}
 u^{k} =\text{prox}_{\tau_{k-1} \eta||\Phi (\cdot)||_0}\big(u^{k-1}-\tau_{k-1}(\mathcal{F}^*(r^{k-1})-D^*(h^{k-1}))\big)\label{u solution}
 \end{align}
 where
 \begin{align}
  \text{prox}_{\tau ||\Phi(\cdot)||_0} (z) =\min_{{u}} \frac{1}{2}||u- z||_2 + \tau||\Phi u||_0 \label{BM3D denoising prob},
 \end{align} 
 which is known as the BM3D denoising problem \cite{BM3D 1}. Complications in (\ref{BM3D denoising prob})
 arise from the fact that $\Phi$ is not an orthonormal transform \cite{BM3D 2} and hence, unlike the case with most X-let transforms, a simple closed-form solution for (\ref{BM3D denoising prob}) does not exist. In \cite{BM3D 1} a heuristic method involving a hard thresholding step followed by  Wiener-filtering was presented for solving (\ref{BM3D denoising prob}). In \cite{BM3D-MRI} the Wiener-filtering was discarded and a faster but approximate solution 
 for (\ref{BM3D denoising prob}) involving only the hard thresholding operation was proposed as
 \begin{equation}\label{solution for BM3D}
u \approx  \Psi\mathcal{H}_{\tau}(\Phi z),
 \end{equation} 
where $\Psi$, called the \textit{BM3D \textit{synthesis} frame}, satisfies $\Psi \Phi = \mathcal{I}$ \cite{BM3D 2} and the hard thresholding operator
$\mathcal{H} $
is defined componentwise by 
\begin{equation}\label{gradian}
\begin{split}
\big(\mathcal{H}_{\tau}(\omega)\big)(i,j) :=  \left\{ \begin{array}{lcr}
     0 , & |\omega(i,j)|  < \sqrt{2 \tau} \\ 
     \omega(i,j), & |\omega(i,j)|\geq \sqrt{2 \tau} \\
    \end{array}\right.,
\end{split}
\end{equation}
for an $\omega \in \text{range}(\Phi)$. Translated to our method, such a solution for (\ref{u solution}) would be
\begin{equation}\label{u sol}
u^k \approx \Psi \mathcal{H}_{\eta\tau_{k-1}} \bigg(   \Phi\big( u^{k-1}-\tau_{k-1}(\mathcal{F}^*(r^{k-1})-D^*(h^{k-1}))  \big)\bigg).
\end{equation}
In order to stay consistent with \cite{BM3D-MRI} and provide fair comparisons with that method, we also choose this solution in lieu of the exact one in \cite{BM3D 1}.

 The updates for $\{v_s:s\in S\}$ are alike and given by shrinkage operation:
\begin{align}
v_s^{k} &= \text{prox}_{\tau_{k-1} \lambda ||\cdot||_{1,2} } (v_s^{k-1} - \tau_{k-1} L_s(h^{k-1})) \notag ,\end{align}
where
\begin{align}\label{v sol} \bigg(\text{prox}_{\alpha ||\cdot||_{1,2}}(v)\bigg)(i,j) &= v(i,j) - \frac{v(i,j)}{\max\{||v(i,j)||_{2},\alpha \}}.
\end{align}
{The $y$-subproblem}:
Similar to the previous step, the  subproblem (\ref{y}) decouples into two stages, one for each variable.

Update for $r$ is given by
\begin{align}
r^{k} =\text{prox}_{\beta \tau_k (\frac{1}{2}||\cdot||_2^2 +\langle \cdot,b \rangle ) }(r^{k-1}+\beta \tau_k \mathcal{F_{\mathcal{M}}}(\bar{u}^k)),
\end{align}  
where 
\begin{align} \label{r sol}
\text{prox}_{\alpha (\frac{1}{2}||\cdot||_2^2 + \langle \cdot , b \rangle)}(r) = \frac{r -\alpha b}{ 1+ \alpha}.
\end{align}

Update for $h$
 is computed via 
\begin{align}
h^{k} = h^{k-1} + \beta \tau_k (-D(\bar{u}^{k}) + \sum_{s\in S}L_s^*(\bar{v}_s^{k})).
\end{align}
The proposed method is summarized in Algorithm \ref{proposed algorithm}.  
 \begin{algorithm} [t]
  \caption{Proposed method for compressed MR image reconstruction}
  \begin{algorithmic}[1]\label{proposed algorithm}
  \renewcommand{\algorithmicrequire}{\textbf{Initialization:}}
  \renewcommand{\algorithmicensure}{\textbf{Output:}}
  \REQUIRE Choose $\theta_0 = 1,\, u^0 = u_\text{zf},\, v_s^0=0$ for $s\in S$, $ h^0=0,\,r^0=0,\, $ $ \tau_0>0,\, \beta>0,\, \mu \in (0,1)$ and $ \delta\in (0,1)$.  
  \\ \textit{\textbf{While convergence criterion not met, repeat:}}
   \STATE \label{step 1} $ \notag
   u^k = \text{prox}_{\tau_{k-1}\eta||\Phi(\cdot)||_0}\big(u^{k-1}-
   \tau_{k-1}(\mathcal{F}_{\mathcal{M}}^*(r^{k-1})-D^*(h^{k-1}))\big);$
   \STATE $v_s^k = \text{prox}_{\tau_{k-1}\lambda ||\cdot||_{1,2}}\big( v_s^{k-1}-\tau_{k-1}(L_s(h^{k-1}))\big),\,\forall s\in S;$
  \STATE Choose $\tau_k \in [\tau_{k-1}, \tau_{k-1}\sqrt{1+\theta_{k-1}}];$
  \\ \textit{\textbf{Linesearch:}}
   \STATE $\theta_k = \frac{\tau_k}{\tau_{k-1}};$ \label{linesearch}
   \STATE $\bar{u}^k = u^k + \theta_k(u^k-u^{k-1});$
   \STATE $\bar{v}^k_s = v_s^k+ \theta_k(v_s^k-v_s^{k-1}),\,\forall s\in S;$
   \STATE $r^{k} = \text{prox}_{\beta \tau_k (\frac{1}{2}||\cdot||_2^2+\langle \cdot,b \rangle)} (r^{k-1} + \beta \tau_k \mathcal{F}_{\mathcal{M}}(\bar{u}^k));$
   \STATE $ h^k  = h^{k-1} + \beta \tau_k (-D(\bar{u}^k)+\sum_{s \in S}L_s^*(\bar{v_s}^k));$
   \IF {$\sqrt{\beta}\tau_k\Big\Vert  \big(\mathcal{F}^*_{\mathcal{M}}(r^{k}-r^{k-1}) - D^*(h^k-h^{k-1}), (L_s (h^k-h^{k-1}))_{s\in S}\big)\Big\Vert_2 $ $\leq \delta \big\Vert(r^{k}-r^{k-1},h^{k}-h^{k-1})\big\Vert_2  $}
   \STATE Return to step \ref{step 1} (break linesearch),
   \ELSE
   \STATE Set $\tau_k = \mu \tau_k$ and return to step \ref{linesearch} (apply another iteration of linesearch).
   \ENDIF
  \ENSURE  Reconstructed MR image $u$, solution to (\ref{proposed model}).
  \end{algorithmic} 
  \end{algorithm}
  We remark that the convergence proof given in \cite{MP} requires that $g$ be convex, however (\ref{g}) is non-convex. If we were to remain strictly within boundaries of convex analysis, the sensible option would be to use $||\Phi u||_1$ instead of $||\Phi u||_0$. However, $l_0$-norm naturally produces sparser representations resulting in much better reconstructions and in practice, whenever $g$ is non-convex but $\text{prox}_g$ is well-defined (as is the case with $l_0$-norm), the convergence behavior of the algorithm remains intact (although we are yet to prove this theoretically). For these reasons we prefer the formulation (\ref{proposed model}). A similar observation was carried out in \cite{FDLCP} where $l_0$-norm regularization was preferred to $l_1$-norm. Furthermore, as we mentioned above another possible framework for solving (\ref{proposed model}) includes ADMM-type methods which are highly popular for constrained problems. In Section \ref{sec exp} we provide numerical experiments that compare the two frameworks.

\section{Results}\label{sec exp}
\subsection{Experiment setup}\label{sec setup}
In this section, we demonstrate the performance of the proposed method on some test data.
All simulations were conducted in MATLAB R2016a on a PC with an AMD FX-7600 Radeon R7 CPU at 2.70 GHz clock speed, AMD R9 M280X GPU with 4GB of memory and 8GB of RAM. 
The parameters in (\ref{proposed model}) and Algorithm \ref{proposed algorithm} are fixed at 
 $\eta=0.2$, $\lambda=\nicefrac{10^{-3}}{7}$, $\mu = 0.7$, $\delta=0.99$, $\beta = 0.016$ and $\tau_0 = \nicefrac{8}{7}$. In the implementation of BM3D frames $\Phi$ and $\Psi$ we used the default choices recommended by the authors of \cite{BM3D 1,BM3D 2} which were also used in \cite{BM3D-MRI}.
We quantify the quality of reconstructed images based on SNR, SSIM and HFEN indices. SNR and SSIM are taken respectively from \cite{TGVSH} and \cite{FDLCP}. { HFEN, originally defined in \cite{DLMRI},  
is implemented by 
     $$ \text{HFEN}(u,u_0) := \frac{||\text{LoG}(u)-\text{LoG}(u_0)||_2}{||\text{LoG}(u_0)||_2} $$
    where $u_0$ is the ground truth reference and $\text{LoG}$ is a rotationally symmetric Laplacian of Gaussian filter with a kernel size of $15\times 15$ pixels and standard deviation of 1.5 pixels.}
  We note that the perfect reconstruction has $\text{SNR} = \infty$, $\text{SSIM} = 1$ and HFEN $=0$.
 \begin{figure}\centering
  \includegraphics[width=0.7\linewidth]{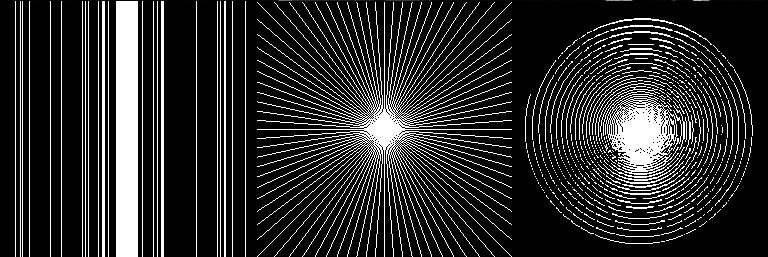}\caption{Sampling patterns used in the experiments of this paper.} \label{data set}
  \end{figure}
  
  {In order to validate our method we collected 100 MR images, including a knee collection of 50 slices from the TSE datasets in \cite{NYU}, and a brain collection of 50 images consisting of 15 $T_2$-weighted slices from \cite{T2}, 15 $T_1$-weighted MPRAGE slices from \cite{T1}, 15 diffusion tensor slices from \cite{DTI} and 5 slices from published papers \cite{LDP,DLMRI,Deep ADMM} in both axial and sagittal views. This diversity in contrast was specifically chosen to demonstrate the versatility of the proposed framework.   }

 
 \subsection{ADMM-based implementation} \label{ADMM}
 ADMM-type algorithms are popular in imaging applications and in particular in compressed MR imaging. For instance, the models proposed in \cite{TGVSH, FDLCP, GBRWT} are all solved via different variants of ADMM and  the deep architecture of \cite{Deep ADMM} is based on this algorithm. The Chambolle-Pock method which was used in \cite{TGV} can also be viewed as a pre-conditioned version of ADMM \cite{Cham-Pock}.  
 
 In this subsection we investigate how such an algorithm would perform in comparison to Algorithm \ref{proposed algorithm}. Direct application of conventional ADMM iterations to (\ref{proposed model}) would lead to subproblems that are actually harder to solve than (\ref{proposed model}) itself. Instead, we consider \textit{generalized alternating direction method of multipliers} (GADMM) \cite{GADMM} which reduces these nearly-intractable subproblems to proximal operators. A particular case of GADMM applied to (\ref{proposed model}) is summarized in Algorithm \ref{GADMM algorithm} (see also Algorithm 3 in \cite{Condat}).
 
 \begin{algorithm} 
   \caption{GADMM for solving (\ref{proposed model})}
   \begin{algorithmic}[1]\label{GADMM algorithm}
   \renewcommand{\algorithmicrequire}{\textbf{Initialization:}}
   \renewcommand{\algorithmicensure}{\textbf{Output:}}
   \REQUIRE Choose $\mu>0$, $\tau < \frac{1}{||D||+\mu}$, $\gamma < \frac{1}{\sum_s ||L_s||}$ and set $u^0 = u_{\text{zf}}$, $v_s^0 = 0$ for $s\in S$ and $\xi^0=0$.  
   \\ \textit{\textbf{While convergence criterion not met, repeat:}}
    \STATE $u^{k+1}=\text{prox}_{\tau \mu \eta ||\Phi(\cdot)||_0} \big( u^k - \tau D^*(Du^k - \sum_{s\in S}L_s^*(v_s^k) +\mu \xi^k ) - \tau \mu \mathcal{F}_{\mathcal{M}}^*(\mathcal{F}_{\mathcal{M}}u^k-b  )\big);$
    \STATE $v_s^{k+1} = \text{prox}_{\gamma\mu\lambda ||\cdot||_{1,2}} \big( v_s^k + L_s (Du^{k+1} - \sum_{s\in S}L_s^*(v_s^k) + \mu \xi^k) \big),\,$ $\forall s\in S; $
    \STATE $\xi^{k+1} = \xi^k + \frac{1}{\mu} (Du^{k+1} - \sum_{s\in S} L_s^*(v_s^{k+1}) ); $
   \ENSURE  Reconstructed MR image $u$, solution to (\ref{proposed model}).
   \end{algorithmic} 
   \end{algorithm}
 
 To test the  performance of GADMM we choose a Shepp-Logan phantom of size $256 \times 256$ and simulate the k-space data by taking the FFT of the phantom. Then we sample the k-space along 12 radial projections and reconstruct with Algorithm \ref{proposed algorithm} and Algorithm \ref{GADMM algorithm}. Parameter choices for Algorithm \ref{proposed algorithm} were given in Subsection \ref{sec setup}. In Algorithm \ref{GADMM algorithm} we set $\tau = \nicefrac{1}{8+\mu}$, $\gamma = \nicefrac{1}{4}$ and tune $\mu$ for optimal performance. 
 
 \begin{figure}\centering
 \includegraphics[width = 0.9\linewidth]{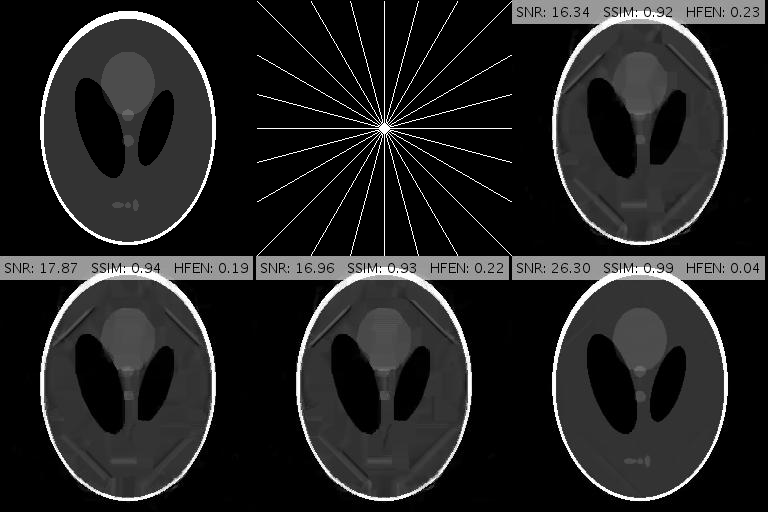}
 \caption{Reconstruction of a  Shepp-Logan phantom from 12 radial spokes via GADMM and Malitsky-Pock methods. Top left: ground truth. Top middle: sampling pattern. Top right: GADMM solution with $\mu = 10^2$. Bottom left: GADMM solution with $\mu = 10^4$. Bottom middle: GADMM solution with $\mu=10^6$. Bottom right: solution by Algorithm \ref{proposed algorithm}. }
  \label{GADMM results}
 \end{figure}
 \begin{figure}
\includegraphics[width = 0.49\linewidth]{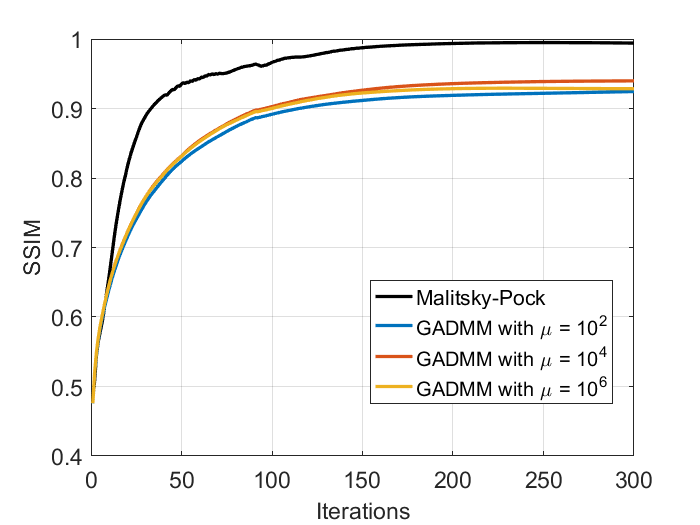}
 \includegraphics[width=0.49\linewidth]{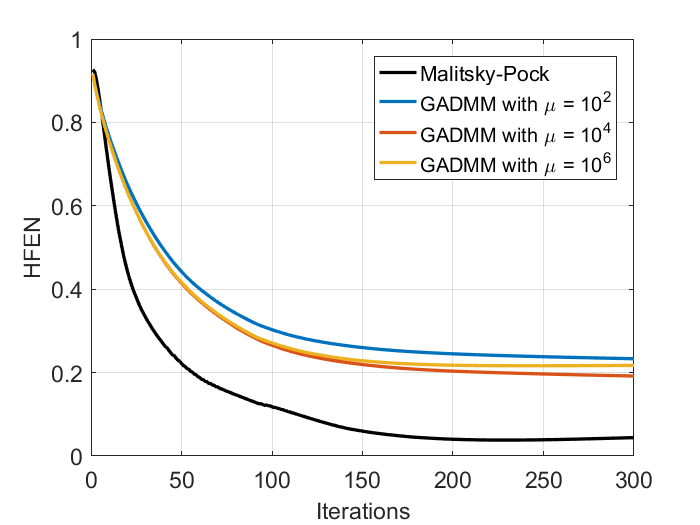}
 \caption{SSIM (left) and HFEN (right) plots for reconstructions in Fig. \ref{GADMM results}.}
 \label{GADMM plots}
 \end{figure}
  
{
 Fig. \ref{GADMM results} compares reconstruction results for GADMM and Algorithm \ref{proposed algorithm} after 300 iterations of each method. Algorithm \ref{proposed algorithm} outperforms GADMM in SNR by more than 8dB and removes almost all the artifacts that GADMM fails to eliminate. SSIM and HFEN plots for these methods are presented in Fig. \ref{GADMM plots}. The three variants of GADMM run for 121.2 ($\mu=10^2$), 115.1 ($\mu=10^4$) and 110.2 ($\mu=10^6$) seconds while the Malitsky-Pock method runs for 119.6 seconds.} We remark that numerical results of GADMM in this section were only provided to justify our unusual choice of Malitsky-Pock algorithm over ADMM-based methods which are more popular in constrained minimization settings. In the remainder of the paper we solely focus on Algorithm \ref{proposed algorithm} and discard GADMM from further consideration.

{ \subsection{Isotropic property of RITV} \label{TV, TGV and RITV isotropy}  }

 \begin{figure}[]\centering
    \minipage{0.45\linewidth}
      \includegraphics[width=\linewidth]{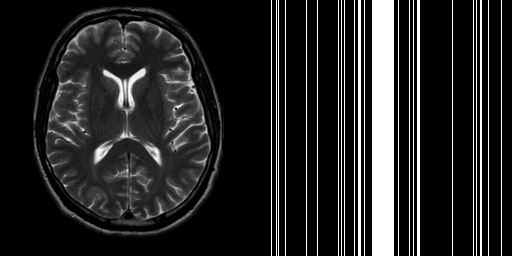}\\
      \centering \footnotesize (1)
    \endminipage \hspace{18pt}
    \minipage{0.45\linewidth}
      \includegraphics[width=\linewidth]{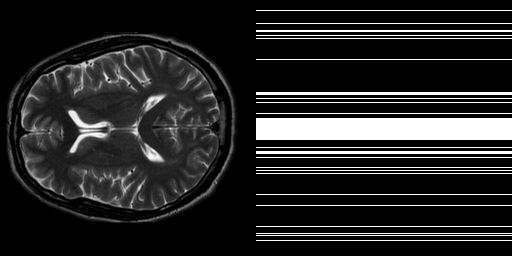}
    \\ \centering \footnotesize  (2) 
    \endminipage
    \normalsize\caption{Simulated upright (1) and rotated (2) imaging setups.}
    \label{rotation_setup}
    \end{figure}
  
      \begin{figure}[h]\centering
      \begin{tikzpicture} \begin{scope}[] \node[anchor=north west,inner sep=0] (image) at (0,0) { \begin{tabular}{l r} \rotatebox[origin=c]{90}{\textbf{TV}}  &   \includegraphics[scale=0.28 ,valign = m]{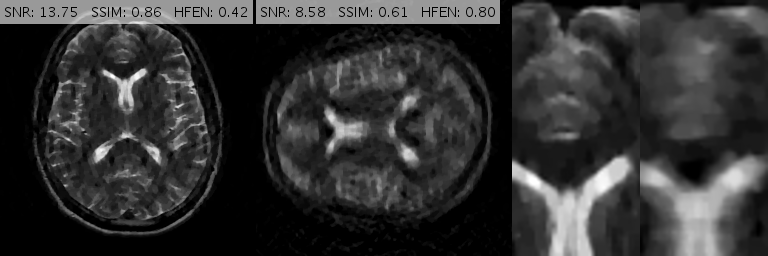} \end{tabular}}; \end{scope} \end{tikzpicture}%
      
      \begin{tikzpicture} \begin{scope}[] \node[anchor=north west,inner sep=0] (image) at (0,0) { \begin{tabular}{l r} \rotatebox[origin=c]{90}{\textbf{TGV}}  &   \includegraphics[scale=0.28 ,valign = m]{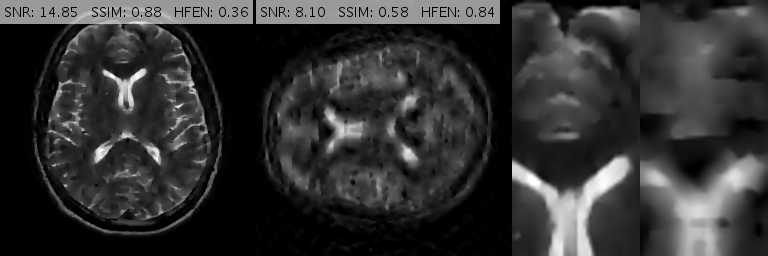} \end{tabular}}; \end{scope} \end{tikzpicture}%
      
      \begin{tikzpicture} \begin{scope}[] \node[anchor=north west,inner sep=0] (image) at (0,0) { \begin{tabular}{l r} \rotatebox[origin=c]{90}{\textbf{RITV}}  &   \includegraphics[scale=0.28 ,valign = m]{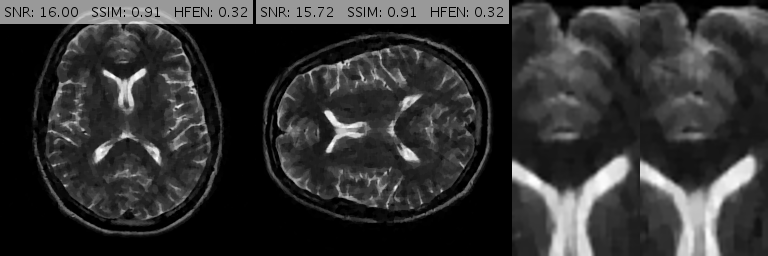} \end{tabular}}; \end{scope} \end{tikzpicture}%
      \caption{Reconstruction of a $T_2$-weighted axial brain image from 20\% Cartesian sampling with TV, TGV and RITV terms. Left column: solution in upright orientation. Middle column: solution in rotated orientation with $\theta=\ang{90}$. Right column: magnified views of upright (left) and rotated (right) reconstructions. Note that the magnified views of  rotated reconstructions have been put in upward position for ease of comparison. }
      \label{rotated_TV_TGV_RITV}
      \end{figure}

 In clinical imaging, the same slice might be acquired in different rotational orientations. For instance, a slice might be imaged in an upright position in one acquisition, but the same slice might be acquired in a rotated position in another experiment. It is ideal to have a robust regularization term that reconstructs the rotated image as if it were acquired in the upward orientation, with no loss of diagnostic value.  The rotation-invariance (or isotropy) of RITV was theoretically established in Theorem \ref{theorem}. In this subsection we investigate the implication of that result in a rotated imaging setup.
 
  We select a $256\times256$-sized $T_2$-weighted axial brain scan from our collection along with a 20\% Cartesian mask. In order to simulate upright and rotated image acquisition sequences, we consider two setups:
 \begin{enumerate}
 \item
 The reference image is in upright orientation. The FFT of the image is taken, then the sampling mask is applied (see Fig. \ref{rotation_setup} (1)).
 \item The reference image is rotated from upright position, say, by $\theta = \ang{90}$ (counterclockwise). The FFT is taken, then the sampling mask rotated by the same $\theta$ is applied (see Fig. \ref{rotation_setup} (2)).   
 \end{enumerate}
 
 \begin{table*}[t]\centering
   \caption{Summary of Experiments}
   \small\addtolength{\tabcolsep}{-4pt}
   \fontsize{7.5}{9}\selectfont
   \begin{tabular}{lcccccccccccccc}
       \toprule
       Sampling  &
       \multicolumn{4}{c}{10\%}
   &&    \multicolumn{4}{c}{20\%}
    &&   \multicolumn{4}{c}{30\%}
    \\ \cline{2-5} \cline{7-10} \cline{12-15} 
    Data &\multicolumn{2}{c}{Brain}
    & \multicolumn{2}{c}{Knee}
    &&\multicolumn{2}{c}{Brain}
        &   \multicolumn{2}{c}{Knee}
     &&\multicolumn{2}{c}{Brain}
         &   \multicolumn{2}{c}{Knee}
     \\ Metric 
      &\multicolumn{1}{c}{SNR}
      &   \multicolumn{1}{c}{SSIM}
      &\multicolumn{1}{c}{SNR}
      &   \multicolumn{1}{c}{SSIM}
      &&\multicolumn{1}{c}{SNR}
      &   \multicolumn{1}{c}{SSIM}
      &\multicolumn{1}{c}{SNR}
      &   \multicolumn{1}{c}{SSIM}
      &&\multicolumn{1}{c}{SNR}
      &   \multicolumn{1}{c}{SSIM}
      &\multicolumn{1}{c}{SNR}
      &   \multicolumn{1}{c}{SSIM}
     \\  \hline
        pFISTA & 14.00$\pm$2.65 & 0.737$\pm$0.05 & 16.63$\pm$1.44 & 0.765$\pm$0.05 && 19.60$\pm$2.65 & 0.902$\pm$0.02 & 23.40$\pm$2.03 & 0.922$\pm$0.02 && 23.43$\pm$2.41 & 0.950$\pm$0.01 & 27.27$\pm$2.20 & 0.963$\pm$0.01 \\
        TGV+Shearlet & 15.00$\pm$2.78 & 0.763$\pm$0.05 & 16.79$\pm$1.56 & 0.729$\pm$0.05 && 20.65$\pm$2.73 & 0.904$\pm$0.02 & 23.43$\pm$2.11 & 0.899$\pm$0.02 && 24.67$\pm$2.51 & 0.952$\pm$0.01 & 27.67$\pm$2.29 & 0.959$\pm$0.01 \\
        GBRWT& 15.90$\pm$2.85 & 0.806$\pm$0.05 & 18.90$\pm$1.82 & 0.826$\pm$0.04 && 22.20$\pm$2.38 & 0.937$\pm$0.02 &         25.49$\pm$2.34 & 0.954$\pm$0.02 && 25.64$\pm$2.07 & 0.967$\pm$0.01 & 28.86$\pm$2.38 & 0.973$\pm$0.01 \\
        FDLCP & 16.51$\pm$3.02 & 0.829$\pm$0.04 & 20.18$\pm$1.99 & 0.857$\pm$0.04 && 22.35$\pm$2.32 & 0.933$\pm$0.02 & 26.20$\pm$2.29 & 0.947$\pm$0.01 && 25.39$\pm$2.06 & 0.962$\pm$0.01 & 29.01$\pm$2.38 & 0.970$\pm$0.01\\
        TL & 15.07$\pm$3.07 & 0.777$\pm$0.06 & 18.35$\pm$1.64 & 0.818$\pm$0.05 && 21.00$\pm$3.01 & 0.915$\pm$0.02 & 24.97$\pm$2.31 & 0.936$\pm$0.02 && 24.07$\pm$2.92 & 0.949$\pm$0.02 & 27.78$\pm$2.42 & 0.964$\pm$0.01 \\
        DAMP & 17.13$\pm$3.14 & 0.847$\pm$0.04 & 20.82$\pm$2.17 & 0.873$\pm$0.04 && 22.28$\pm$2.69 & 0.940$\pm$0.02 & 25.97$\pm$2.56 & 0.952$\pm$0.01 && 25.18$\pm$2.29 & 0.965$\pm$0.01 & 28.93$\pm$2.69 & 0.974$\pm$0.01 \\
        BM3D-MRI& 16.73$\pm$3.01 & 0.845$\pm$0.04 & 20.27$\pm$2.15 & 0.865$\pm$0.04 && 22.39$\pm$2.37 & 0.940$\pm$0.02 & 26.13$\pm$2.57 & 0.953$\pm$0.02 && 25.06$\pm$2.10 & 0.966$\pm$0.01 & 28.89$\pm$2.70 & 0.976$\pm$0.01 \\
        Proposed & \textbf{17.67$\pm$3.05} & \textbf{0.863$\pm$0.03}& \textbf{21.43$\pm$2.16} & \textbf{0.887$\pm$0.03} && \textbf{23.95$\pm$2.45} & \textbf{0.951$\pm$0.02} & \textbf{27.75$\pm$2.62} & \textbf{0.962$\pm$0.07} && \textbf{27.32$\pm$2.09} & \textbf{0.974$\pm$0.01} & \textbf{30.95$\pm$2.80} & \textbf{0.981$\pm$0.00}\\
        \bottomrule
        \label{noiselessTable}
     \end{tabular}
   \end{table*}
   \normalsize
 
  We investigate the performance of RITV compared to TV and TGV in these setups. The implementation of RITV is the same as described in Algorithm \ref{proposed algorithm} and Subsection \ref{sec setup}, with the only difference being the removal of the BM3D term (and readjustment of $\beta$ to $1.7\times 10^{-5}$), since we wish to observe the performance of RITV only. TGV is implemented using the algorithm and parameters described in \cite{TGV MRI}. TV is also implemented with the same algorithm, with $\alpha_1$ optimized to $ 10^{-3}$. TV and TGV algorithms run for 500 iterations while the RITV runs for 200.  
 
 Fig. \ref{rotated_TV_TGV_RITV} depicts the reconstruction results for above setups. As attested by the magnified views and error metrics inserted to the top of each reconstruction, TV and TGV significantly degrade the reconstructed image in the rotated setup while RITV gives a remarkably consistent and high-quality result which is hardly distinguishable from the upright solution (note that HFEN and SSIM indices do not change in RITV after rotation). Of course, the \textit{upright} and \textit{rotated} directions are only chosen relatively and can always be swapped, nevertheless, the point is  TV and TGV fail in at least one of the orientations while RITV succeeds in both. Note that if the selected MR image and Cartesian mask are denoted by $u$ and $\mathcal{M}$ respectively, then a simple machine-aided computation shows that $28.01=||\mathcal{M}\odot u||_2\neq ||\mathcal{M}\odot(\mathcal{R}u)||_2 = 24.42$. Therefore, even though the norm-preserving assumption of Theorem \ref{theorem} does not hold true in this experiment (and in fact in many practical situations), the result of RITV is still incredibly isotropic. 
 
 \subsection{Comparison with other works: noise-free scenario} \label{other works}

  We compare the proposed method with various state-of-the-art algorithms including \textbf{TGV+Shearlet} \cite{TGVSH}, \textbf{pFISTA} \cite{pFISTA}, \textbf{Transform Learning (TL)} \cite{TLMRI},\textbf{ Deep ADMM Network} \cite{Deep ADMM},\textbf{ BM3D-MRI} \cite{BM3D-MRI}, \textbf{Graph Based Redundant Wavelet Transform (GBRWT)} \cite{GBRWT}, \textbf{Fast Dictionary Learning on Classified Patches (FDLCP)} \cite{FDLCP} and {\textbf{Denoising Approximate Message Passing (DAMP)} \cite{DAMP}}. Some other methods known to the community include PANO \cite{PANO}, PBDW \cite{PBDW} and FCSA \cite{FCSA}, however, these methods are known to be outperformed by the methods we mentioned above and hence have not been included in our comparisons to save space. Furthermore, a more recent version of TL is proposed in \cite{FRIST} called FRIST. This method outperforms TL by an average SNR of 0.3 dB \cite{FRIST} while demanding much higher computational cost, and hence has not been included in our experiments. In our comparisons we used the software packages graciously provided by the authors of these works. We kept the recommended default settings in all of them, unless declared otherwise. MATLAB codes for the proposed method will also be published in \cite{Dataset}.  {We remark that the primary focus of the presented paper is on the noiseless case since this scenario lays the groundwork for all MRI reconstruction algorithms. Most of the algorithms mentioned above also focus on this scenario alone. The noisy setting only requires a readjustment in parameters and does not pose a new challenge. Hence, investigation of the proposed algorithm in noisy scenario is postponed to Subsection \ref{noisy scenario}.}

 \begin{table}\centering \caption {Approximate Computation Times ($\pm3$ minutes) in Table \ref{noiselessTable}.} \begin{tabular}{ll|ll} \toprule Method & Time (min) & Method & Time (min) \\ \hline TGV+Shearlet & 65 & TL & 260 \\ pFISTA & 25 & GBRWT & 300 \\  FDLCP & 320 & BM3D-MRI & 105 \\ Proposed  & 50 & DAMP & 60 \\ \bottomrule   \label{Table1} \end{tabular} \end{table} 
 
{
Table \ref{noiselessTable} presents a summary of the experiments in this subsection. The spiral trajectory was chosen in Table \ref{noiselessTable} (similar results were seen with other patterns but not presented to save space). For each sampling rate and each dataset (brain, knee), 50 experiments (involving 50 slices as described in Subsection \ref{sec setup}) were conducted, and mean $\pm$ standard deviation values were inserted into the table. Hence, a total of 2400 experiments are reported in Table \ref{noiselessTable}. All slices were mapped to the interval $[0,1]$ and resized to $128 \times 128$ for computational practicality  (a summary of computation times is also given in Table \ref{Table1}). 
In Table \ref{noiselessTable} all methods (except for FDLCP and GBRWT) ran for 100 iterations which was sufficient for their convergence. 
We remark that the authors of FDLCP, GBRWT and DAMP have shared an obfuscated (non-open source) version of their codes. Consequently, we could not observe the SNR, SSIM and HFEN plots for these methods in Figs. \ref{sag SNR plot}, \ref{knee SNR plot} and \ref{axial HFEN plot}. Furthermore, GBRWT and FDLCP use an internally-defined stopping criterion.

\begin{figure} \includegraphics[height=4.5cm,width=0.495\linewidth]{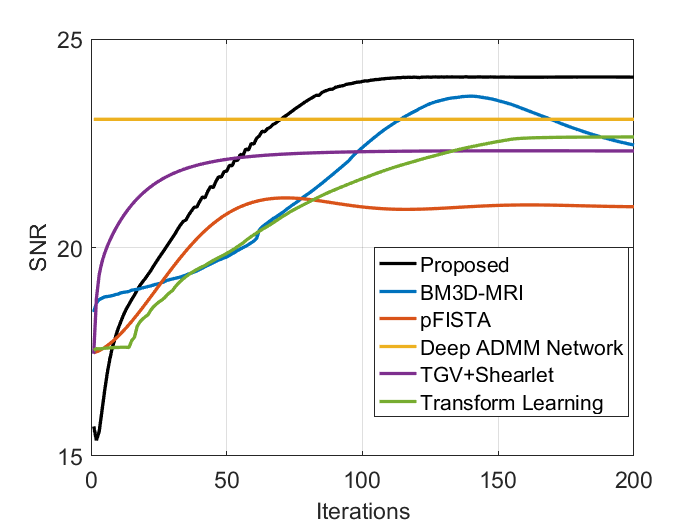}\hspace{0pt}  \includegraphics[height=4.5cm,width=0.495\linewidth]{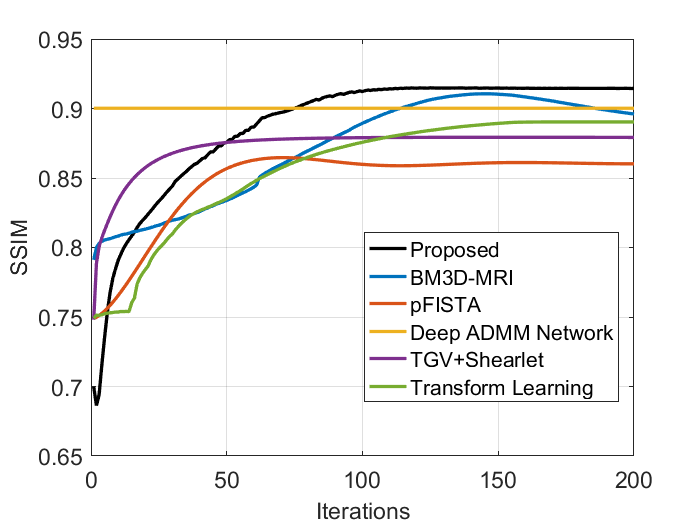}\hspace{0pt} \caption{SNR (left) and SSIM (right) plots for the reconstructions in Fig. \ref{sag}.} \label{sag SNR plot} \end{figure}
       
\begin{figure}  \includegraphics[height=4.5cm,width=0.495\linewidth]{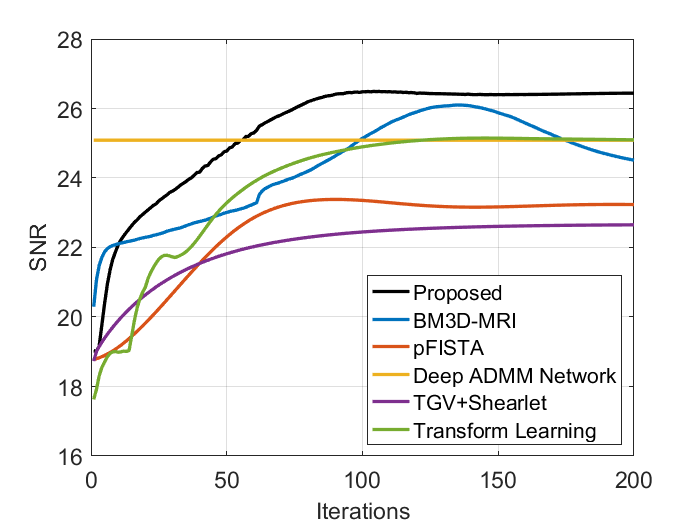}  \includegraphics[height=4.5cm,width=0.495\linewidth]{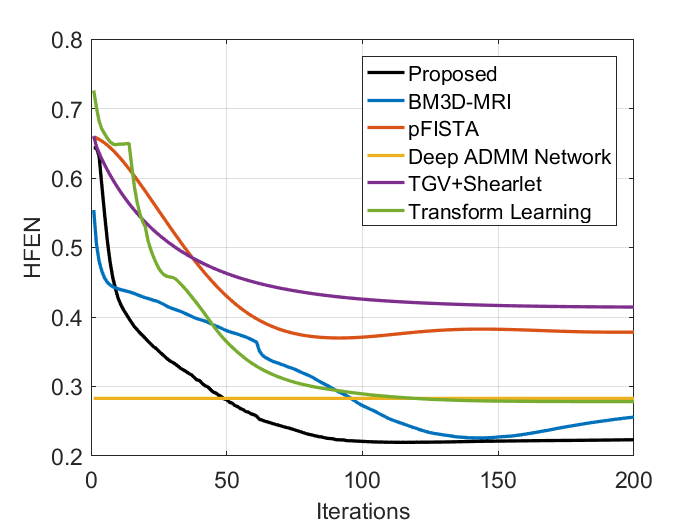}  \caption{SNR (keft) and HFEN (right) plots for reconstructions in Fig. \ref{knee}.} \label{knee SNR plot} \end{figure}
        
\begin{figure}      \includegraphics[height=4.5cm,width=0.495\linewidth]{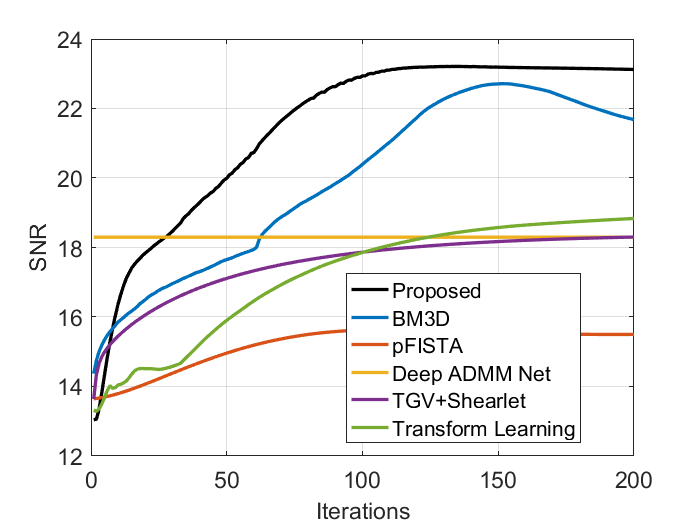}\hspace{0pt}  \includegraphics[height=4.5cm,width=0.495\linewidth]{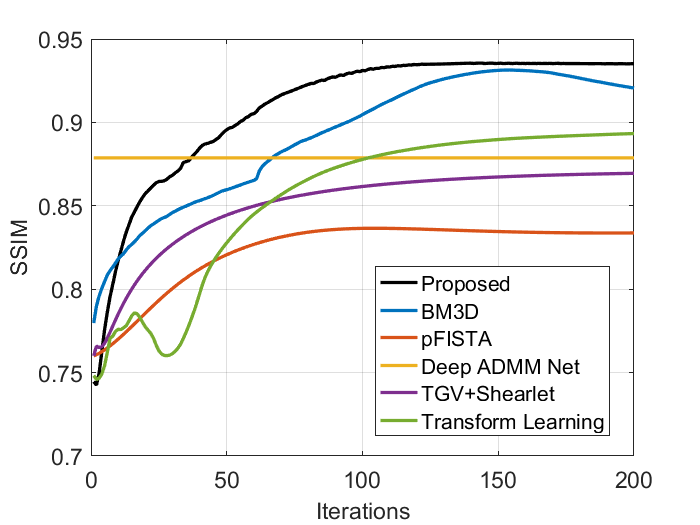}\hspace{0pt}  \caption{SNR (left) and SSIM (right) plots for reconstructions  in Fig. \ref{axial}.}\label{axial HFEN plot}   \end{figure}   
   
All methods are initialized with the zero-filling solution $u_{\text{zf}}$ except for FDLCP and GBRWT which are initialized with a solution provided by a shift-invariant discrete wavelet transform-based (SIDWT) algorithm \cite{GBRWT, FDLCP}.
  
  The FDLCP algorithm comes with an option to choose between convex $||\cdot||_1$ and non-convex $||\cdot||_0$ regularization. We choose $||\cdot||_0$ which is the superior version.
    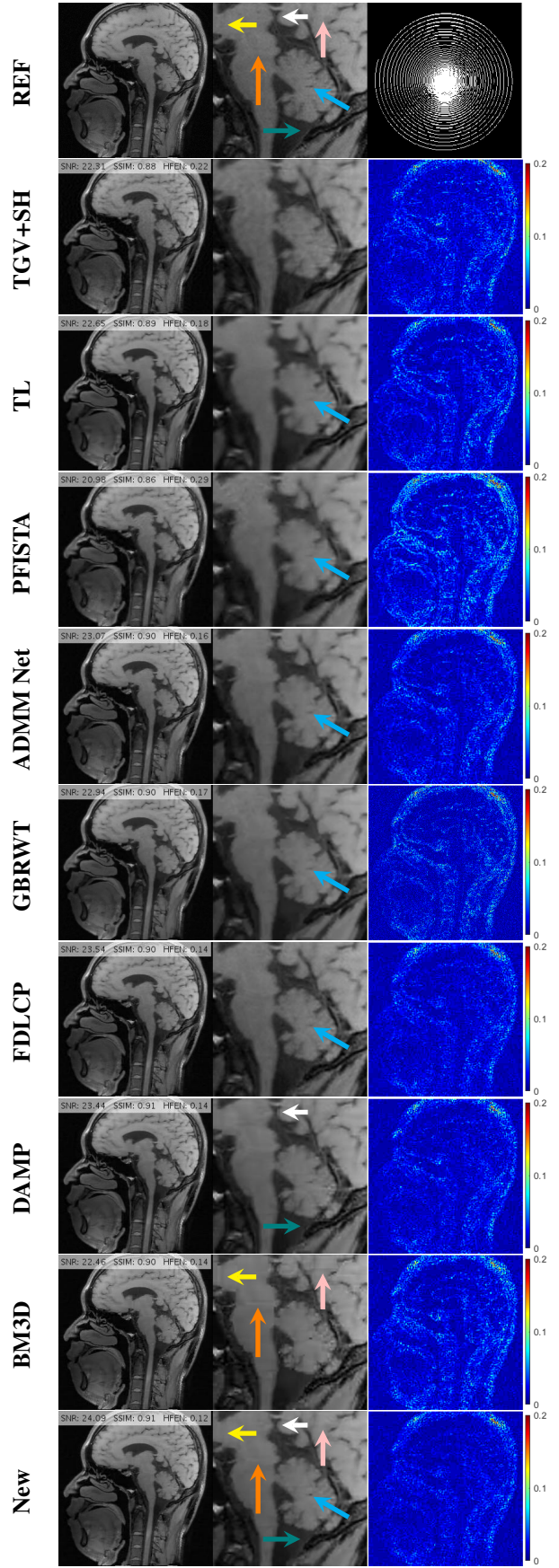
\begin{figure}\centering
    
    \begin{tikzpicture} \node [anchor=north] (water1) at (3.8,-1.5) {}; \node [anchor=north] (water2) at (3.9,-0.2) {}; \node [anchor=north] (water3) at (4.75,-0.8) {}; \node [anchor=north] (water) at (5.25,-1.5) {};\node [anchor=north] (water4) at (3.75,-1.75) {}; \node [anchor=north] (water5) at (4.65,-0.08) {}; \begin{scope}[] \node[anchor=north west,inner sep=0] (image) at (0,0) {  \begin{tabular}{l r} \rotatebox[origin=c]{90}{\textbf{REF}}  &  \includegraphics[scale=0.25 ,valign = m]{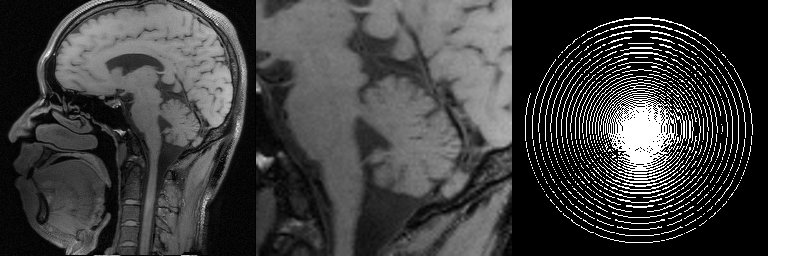}   \end{tabular}}; \draw [-stealth, line width=2pt, cyan] (water) --++ (-0.64,0.37); \draw [-stealth, line width=2pt, orange] (water1) --++ (0,0.85); \draw [-stealth, line width=2pt, yellow] (water2) --++ (-0.55,0);  \draw [-stealth, line width=2pt, pink] (water3) --++ (0,0.65);\draw [-stealth, line width=2pt, teal] (water4) --++ (0.67,0); \draw [-stealth, line width=2pt, white] (water5) --++ (-0.5,0);
     \end{scope}\end{tikzpicture}%
    
    \begin{tikzpicture} \node [anchor=north] (water) at (5.75,-2) {}; \begin{scope}[] \node[anchor=north west,inner sep=0] (image) at (0,0) { \begin{tabular}{l r} \rotatebox[origin=c]{90}{\textbf{TGV+SH}}  &   \includegraphics[scale=0.25 ,valign = m]{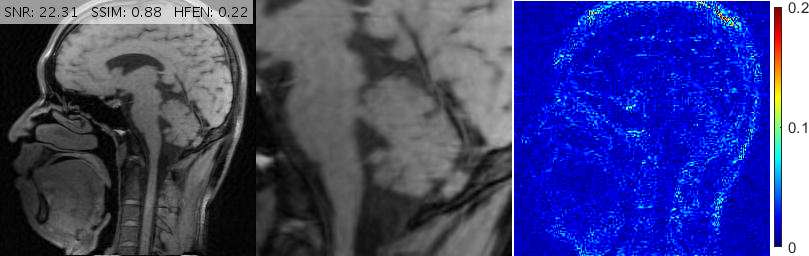} \end{tabular}};  
    \end{scope} \end{tikzpicture}%
    
    \begin{tikzpicture}\node [anchor=north] (water) at (5.25,-1.5) {}; \begin{scope}[] \node[anchor=north west,inner sep=0] (image) at (0,0) { \begin{tabular}{l r} \rotatebox[origin=c]{90}{\textbf{TL}} & \includegraphics[scale=0.25 ,valign = m]{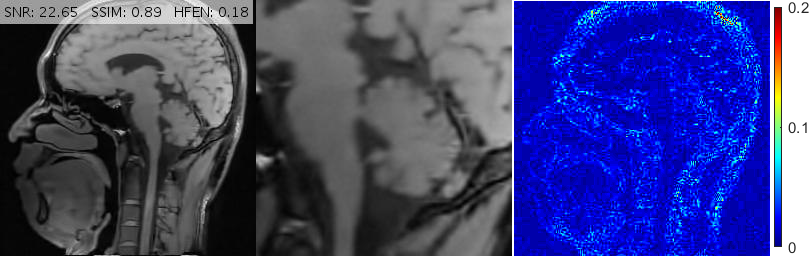} \end{tabular}};  \draw [-stealth, line width=2pt, cyan] (water) --++ (-0.64,0.37); 
    \end{scope} \end{tikzpicture}%
    
    \begin{tikzpicture} \node [anchor=north] (water) at (5.25,-1.5) {}; \begin{scope}[] \node[anchor=north west,inner sep=0] (image) at (0,0) {\begin{tabular}{l r}  \rotatebox[origin=c]{90}{\textbf{PFISTA}}   &  \includegraphics[scale=0.25 ,valign = m]{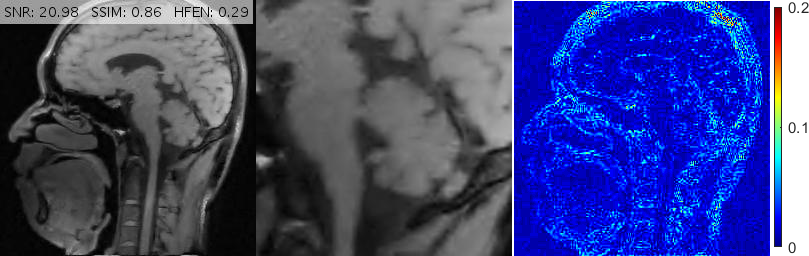} \end{tabular}};  \draw [-stealth, line width=2pt, cyan] (water) --++ (-0.64,0.37); 
    \end{scope} \end{tikzpicture}%
    
    \begin{tikzpicture}\node [anchor=north] (water) at (5.25,-1.5) {}; \begin{scope}[] \node[anchor=north west,inner sep=0] (image) at (0,0) { \begin{tabular}{l r} \rotatebox[origin=c]{90}{\textbf{ADMM Net}} & \includegraphics[scale=0.25 ,valign = m]{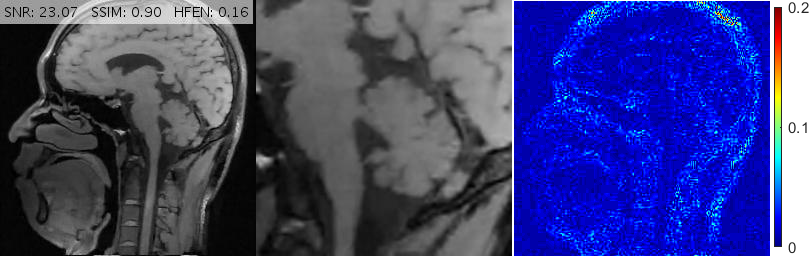} \end{tabular}};  \draw [-stealth, line width=2pt, cyan] (water) --++ (-0.64,0.37); \end{scope} \end{tikzpicture}%
    
    \begin{tikzpicture}
   \node [anchor=north] (water) at (5.25,-1.5) {}; \begin{scope}[] \node[anchor=north west,inner sep=0] (image) at (0,0) { \begin{tabular}{l r} \rotatebox[origin=c]{90}{\textbf{GBRWT}}  &  \includegraphics[scale=0.25 ,valign = m]{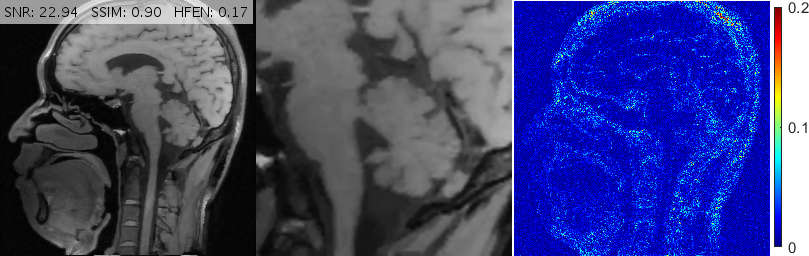} \end{tabular}};  \draw [-stealth, line width=2pt, cyan] (water) --++ (-0.64,0.37); \end{scope} \end{tikzpicture}%
    
    \begin{tikzpicture}
    \node [anchor=north] (water) at (5.25,-1.5) {}; \begin{scope}[] \node[anchor=north west,inner sep=0] (image) at (0,0) {  \begin{tabular}{l r} \rotatebox[origin=c]{90}{\textbf{FDLCP}} & \includegraphics[scale=0.25 ,valign = m]{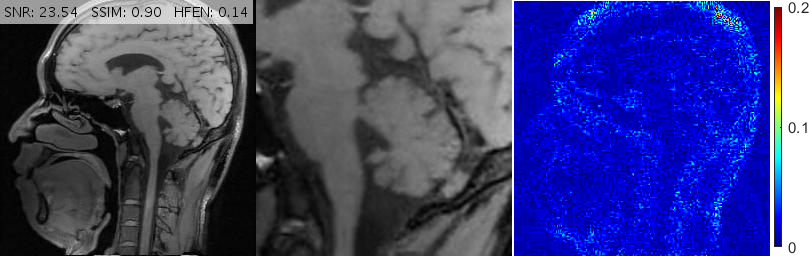} \end{tabular}};  \draw [-stealth, line width=2pt, cyan] (water) --++ (-0.64,0.37); \end{scope} \end{tikzpicture}%
    
    \begin{tikzpicture}
        \node [anchor=north] (water4) at (3.75,-1.75) {}; \node [anchor=north] (water5) at (4.65,-0.08) {};  \begin{scope}[] \node[anchor=north west,inner sep=0] (image) at (0,0) {  \begin{tabular}{cc} \rotatebox[origin=c]{90}{\textbf{DAMP}} & \includegraphics[scale=0.25 ,valign = m]{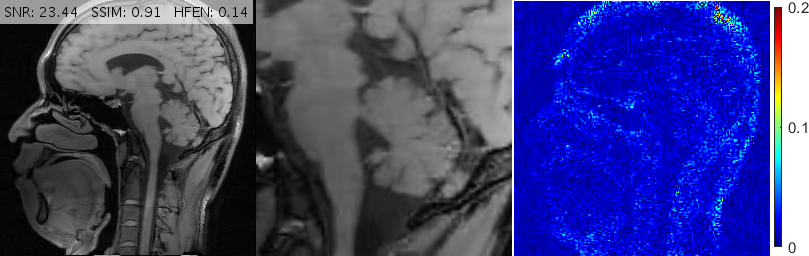} \end{tabular}};  \draw [-stealth, line width=2pt, teal] (water4) --++ (0.67,0); \draw [-stealth, line width=2pt, white] (water5) --++ (-0.5,0);\end{scope} \end{tikzpicture}%
    
    \begin{tikzpicture}
    \node [anchor=north] (water1) at (3.8,-1.5) {}; \node [anchor=north] (water2) at (3.9,-0.2) {}; \node [anchor=north] (water3) at (4.75,-0.8) {};\begin{scope}[] \node[anchor=north west,inner sep=0] (image) at (0,0) { \begin{tabular}{cc} \rotatebox[origin=c]{90}{\textbf{BM3D}}  &  \includegraphics[scale=0.25 ,valign = m]{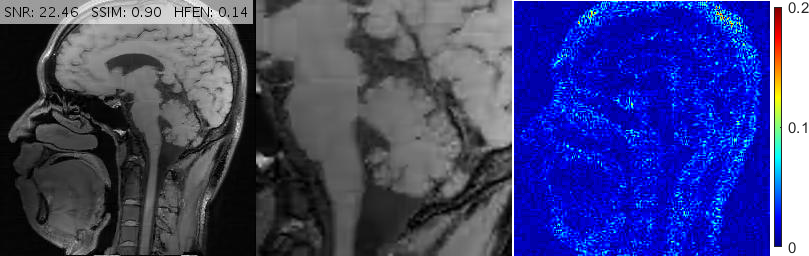}  \end{tabular}};  \draw [-stealth, line width=2pt, orange] (water1) --++ (0,0.85); \draw [-stealth, line width=2pt, yellow] (water2) --++ (-0.55,0);  \draw [-stealth, line width=2pt, pink] (water3) --++ (0,0.65); \end{scope} \end{tikzpicture}%
    
    \begin{tikzpicture} \node [anchor=north] (water1) at (3.8,-1.5) {}; \node [anchor=north] (water2) at (3.9,-0.2) {}; \node [anchor=north] (water3) at (4.75,-0.8) {}; \node [anchor=north] (water) at (5.25,-1.5) {}; \node [anchor=north] (water4) at (3.75,-1.75) {}; \node [anchor=north] (water5) at (4.65,-0.08) {};\begin{scope}[] \node[anchor=north west,inner sep=0] (image) at (0,0) {\begin{tabular}{l r} \rotatebox[origin=c]{90}{\textbf{New}}  &    \includegraphics[scale=0.25 ,valign = m]{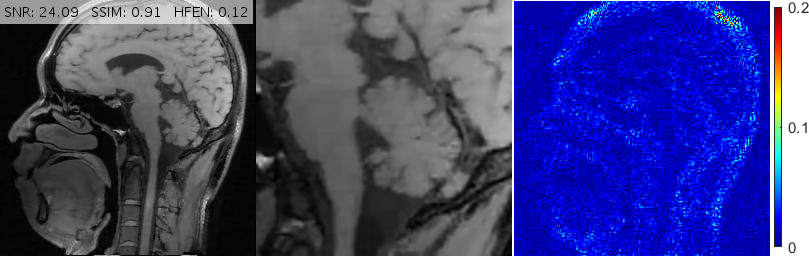}  \end{tabular}}; \draw [-stealth, line width=2pt, cyan] (water) --++ (-0.64,0.37); \draw [-stealth, line width=2pt, orange] (water1) --++ (0,0.85); \draw [-stealth, line width=2pt, yellow] (water2) --++ (-0.55,0);  \draw [-stealth, line width=2pt, pink] (water3) --++ (0,0.65); \draw [-stealth, line width=2pt, teal] (water4) --++ (0.67,0); \draw [-stealth, line width=2pt, white] (water5) --++ (-0.5,0);\end{scope} \end{tikzpicture}%
    
    \caption{Comparison between various reconstructions for a $256\times256$ $T_1$-weighted sagittal head scan with 16\% spiral sampling. The left, middle and right columns respectively show the reconstructions, magnified views and error maps.} \label{sag}
    \end{figure}

     The TL method did not produce acceptable results in our experiments at its default settings. In order to improve the performance of this method we set the sparsity level $s$ to uniformly increase from $0.005$ to $0.165$ in the first 50 iterations and then  fix it at $0.165$ for subsequent iterations. { Furthermore, in accordance with image sizes, $N=128^2$ is used for experiments of Table \ref{noiselessTable} and $N=256^2$ is used in Figs. \ref{sag}, \ref{knee} and \ref{axial}. All other parameters are kept at default settings.} 
     
     {The Deep ADMM Network is not reported in Table \ref{noiselessTable} since it did not produce acceptable results for matrix size $128\times128$ and modifying the architecture to cope with this size appears to be beyond the scope of the current paper. Nevertheless, for matrix size $256\times256$ there is an option to choose between stage-7,  stage-14 and  stage-15 trained networks learned from 100 MR slices. For the experiments in Figs. \ref{sag}, \ref{knee} and \ref{axial} we select the stage-15 network which is the best one.}\footnote{We were also eager to include the deep learning method of \cite{VN} in our tests but we did not receive a response to our request for a MATLAB interface from the first author of that work.}

Fig. \ref{sag} shows the performance of various algorithms for a sagittal head scan reconstruction under 16\% spiral sampling. The solutions provided by pFISTA and TL miss much of the image content due to over-smoothing. TGV+Shearlet leaves many incoherent artifacts on all regions. The magnified views obviously show that BM3D degrades the image by leaving block artifacts around the cerebellum and streaking artifacts on the medulla and the visual cortex. DAMP, GBRWT, ADMM Net and FDLCP provide more accurate results, however, by inspecting the magnified views it becomes evident that some small artifacts are introduced in DAMP near the caudate nucleus and beneath the cerebellum while a dark hole at the center of the cerebellum has been almost entirely smoothed out by the other three methods. The proposed method (labeled `New') accurately captures this hole and corrects all the errors mentioned above.

Fig. \ref{knee} shows reconstructions for an FSE sagittal knee slice under 16\% radial sampling. Solutions provided by TGV+Sh, pFISTA, GBRWT, DAMP and BM3D give rise to incoherent, streaking and block artifacts observable in the magnified views. Deep ADMM Net, TL and FDLCP provide more acceptable results, however, the proposed method gives the best and sharpest reconstruction. 

 \begin{figure}\centering
 \begin{tikzpicture}\node [anchor=north] (water1) at (5.27,-1.79) {}; \node [anchor=north] (water2) at (5.4,-1) {};  \node [anchor=north] (water4) at (3.75,-1.85) {}; \node [anchor=north] (water5) at (3.15,-1.12) {}; \node [anchor=north] (water6) at (4,-0.8) {}; \node [anchor=north] (water7) at (3.4,-0.4) {}; \begin{scope}[] \node[anchor=north west,inner sep=0] (image) at (0,0) {  \begin{tabular}{l r} \rotatebox[origin=c]{90}{\textbf{REF}}  &  \includegraphics[scale=0.25,valign = m]{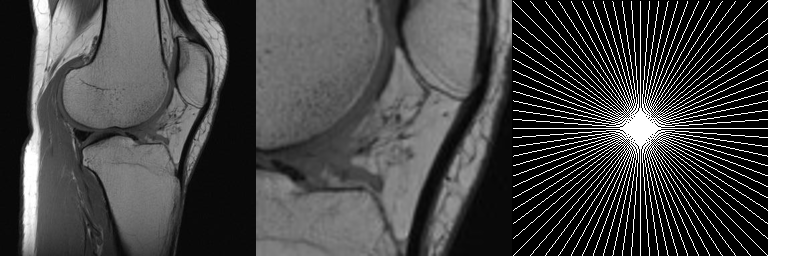}   \end{tabular}}; \draw [-stealth, line width=2pt, cyan] (water1) --++ (-0.68,0.43); \draw [-stealth, line width=2pt, pink] (water2) --++ (-0.68,0); \draw [-stealth, line width=2pt, orange] (water4) --++ (0.6,0); \draw [-stealth, line width=2pt, green] (water5) --++ (0.7,0); \draw [-stealth, line width=2pt, violet] (water6) --++ (-0.6,0); \draw [-stealth, line width=2pt, white] (water7) --++ (0.65,0);
  \end{scope}\end{tikzpicture}%
 
 \begin{tikzpicture} \node [anchor=north] (water1) at (5.27,-1.79) {}; \begin{scope}[] \node[anchor=north west,inner sep=0] (image) at (0,0) { \begin{tabular}{l r} \rotatebox[origin=c]{90}{\textbf{TGV+SH}}  &   \includegraphics[scale=0.25,valign = m]{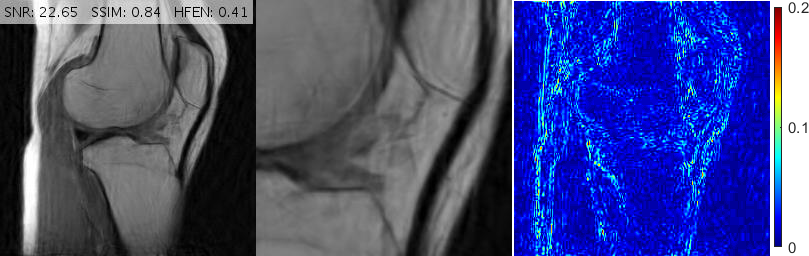} \end{tabular}};  \draw [-stealth, line width=2pt, cyan] (water1) --++ (-0.68,0.43); 
 \end{scope} \end{tikzpicture}%
 
 \begin{tikzpicture}\node [anchor=north] (water1) at (5.27,-1.79) {}; \node [anchor=north] (water4) at (3.75,-1.85) {};\begin{scope}[] \node[anchor=north west,inner sep=0] (image) at (0,0) { \begin{tabular}{l r} \rotatebox[origin=c]{90}{\textbf{TL}} & \includegraphics[scale=0.25,valign = m]{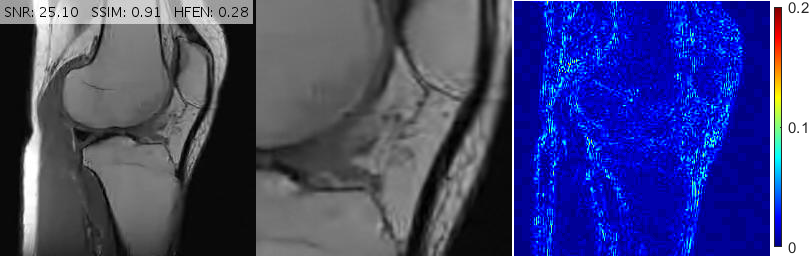} \end{tabular}};  \draw [-stealth, line width=2pt, cyan] (water1) --++ (-0.68,0.43);  \draw [-stealth, line width=2pt, orange] (water4) --++ (0.6,0);
 \end{scope} \end{tikzpicture}%
 
 \begin{tikzpicture}  \begin{scope}[] \node[anchor=north west,inner sep=0] (image) at (0,0) {\begin{tabular}{l r}  \rotatebox[origin=c]{90}{\textbf{PFISTA}}   &  \includegraphics[scale=0.25,valign = m]{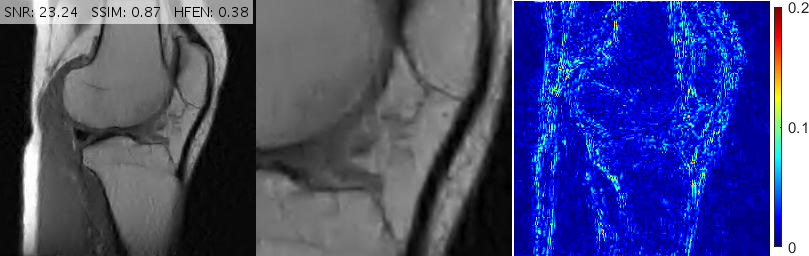} \end{tabular}}; \node [anchor=north] (water1) at (5.27,-1.79) {};
 \draw [-stealth, line width=2pt, cyan] (water1) --++ (-0.68,0.43); 
 \end{scope} \end{tikzpicture}%
 
 \begin{tikzpicture}\node [anchor=north] (water1) at (5.27,-1.79) {};  \begin{scope}[] \node[anchor=north west,inner sep=0] (image) at (0,0) { \begin{tabular}{l r} \rotatebox[origin=c]{90}{\textbf{ADMM Net}} & \includegraphics[scale=0.25,valign = m]{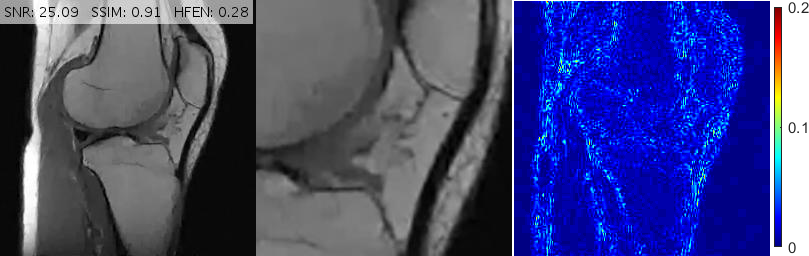} \end{tabular}};  \draw [-stealth, line width=2pt, cyan] (water1) --++ (-0.68,0.43); \end{scope} \end{tikzpicture}%
 
 \begin{tikzpicture}
 \node [anchor=north] (water1) at (5.27,-1.79) {}; \begin{scope}[] \node[anchor=north west,inner sep=0] (image) at (0,0) { \begin{tabular}{l r} \rotatebox[origin=c]{90}{\textbf{GBRWT}}  &  \includegraphics[scale=0.25,valign = m]{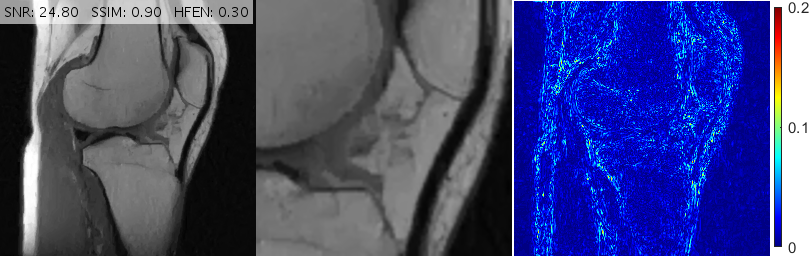} \end{tabular}};  \draw [-stealth, line width=2pt, cyan] (water1) --++ (-0.68,0.43); \end{scope} \end{tikzpicture}%
 
 \begin{tikzpicture}
 \node [anchor=north] (water1) at (5.27,-1.79) {};  \node [anchor=north] (water2) at (5.4,-1) {};  \begin{scope}[] \node[anchor=north west,inner sep=0] (image) at (0,0) {  \begin{tabular}{l r} \rotatebox[origin=c]{90}{\textbf{FDLCP}} & \includegraphics[scale=0.25,valign = m]{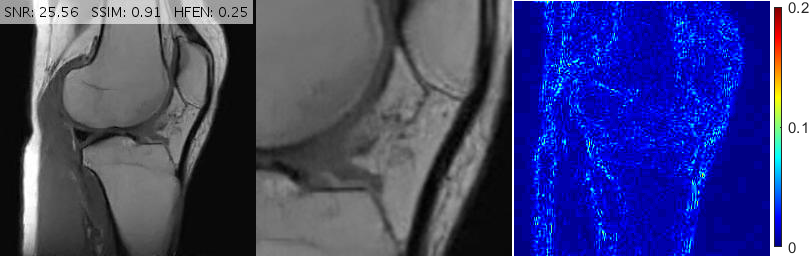} \end{tabular}};  \draw [-stealth, line width=2pt, cyan] (water1) --++ (-0.68,0.43); \draw [-stealth, line width=2pt, pink] (water2) --++ (-0.68,0); \end{scope} \end{tikzpicture}%
 
 \begin{tikzpicture}
  \node [anchor=north] (water6) at (4,-0.8) {}; \node [anchor=north] (water7) at (3.4,-0.4) {}; \begin{scope}[] \node[anchor=north west,inner sep=0] (image) at (0,0) { \begin{tabular}{l r} \rotatebox[origin=c]{90}{\textbf{DAMP}}  &  \includegraphics[scale=0.25,valign = m]{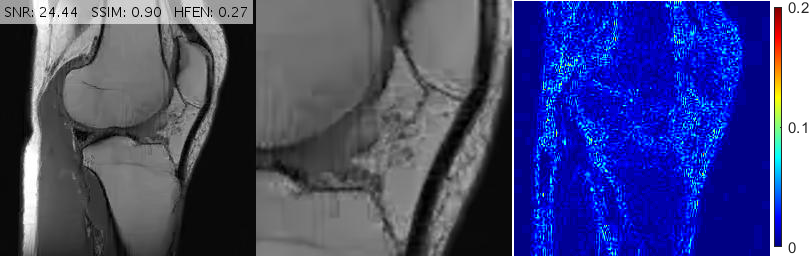}  \end{tabular}};  \draw [-stealth, line width=2pt, violet] (water6) --++ (-0.6,0); \draw [-stealth, line width=2pt, white] (water7) --++ (0.65,0);  \end{scope} \end{tikzpicture}%
 
 \begin{tikzpicture}
 \node [anchor=north] (water5) at (3.15,-1.12) {}; \begin{scope}[] \node[anchor=north west,inner sep=0] (image) at (0,0) { \begin{tabular}{l r} \rotatebox[origin=c]{90}{\textbf{BM3D}}  &  \includegraphics[scale=0.25,valign = m]{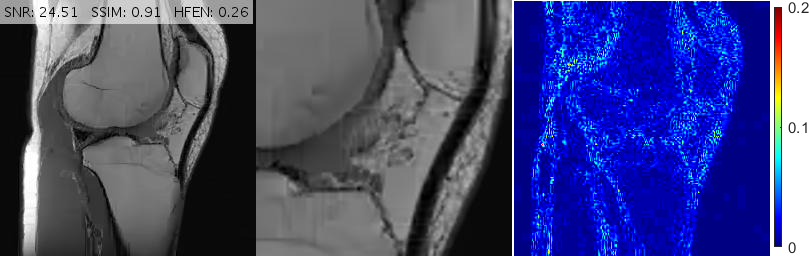}  \end{tabular}};  \draw [-stealth, line width=2pt, green] (water5) --++ (0.7,0);  \end{scope} \end{tikzpicture}%
 
 \begin{tikzpicture} \node [anchor=north] (water1) at (5.27,-1.79) {}; \node [anchor=north] (water2) at (5.4,-1) {};  \node [anchor=north] (water4) at (3.75,-1.85) {}; \node [anchor=north] (water5) at (3.15,-1.12) {}; \node [anchor=north] (water6) at (4,-0.8) {}; \node [anchor=north] (water7) at (3.4,-0.4) {};\begin{scope}[] \node[anchor=north west,inner sep=0] (image) at (0,0) {\begin{tabular}{l r} \rotatebox[origin=c]{90}{\textbf{New}}  &    \includegraphics[scale=0.25,valign = m]{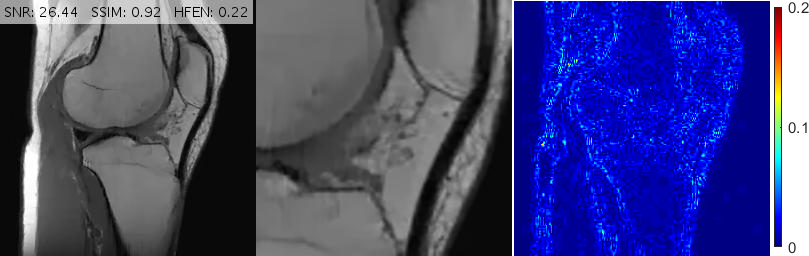}  \end{tabular}};  \draw [-stealth, line width=2pt, cyan] (water1) --++ (-0.68,0.43); \draw [-stealth, line width=2pt, pink] (water2) --++ (-0.68,0); \draw [-stealth, line width=2pt, orange] (water4) --++ (0.6,0); \draw [-stealth, line width=2pt, green] (water5) --++ (0.7,0); \draw [-stealth, line width=2pt, violet] (water6) --++ (-0.6,0); \draw [-stealth, line width=2pt, white] (water7) --++ (0.65,0); \end{scope} \end{tikzpicture}%
 \caption{Comparison between various reconstructions for a $256\times256$ sagittal TSE knee image with 16\% radial sampling.  The left, middle and right columns respectively show the reconstructions, magnified views and error maps.} \label{knee}
 \end{figure}
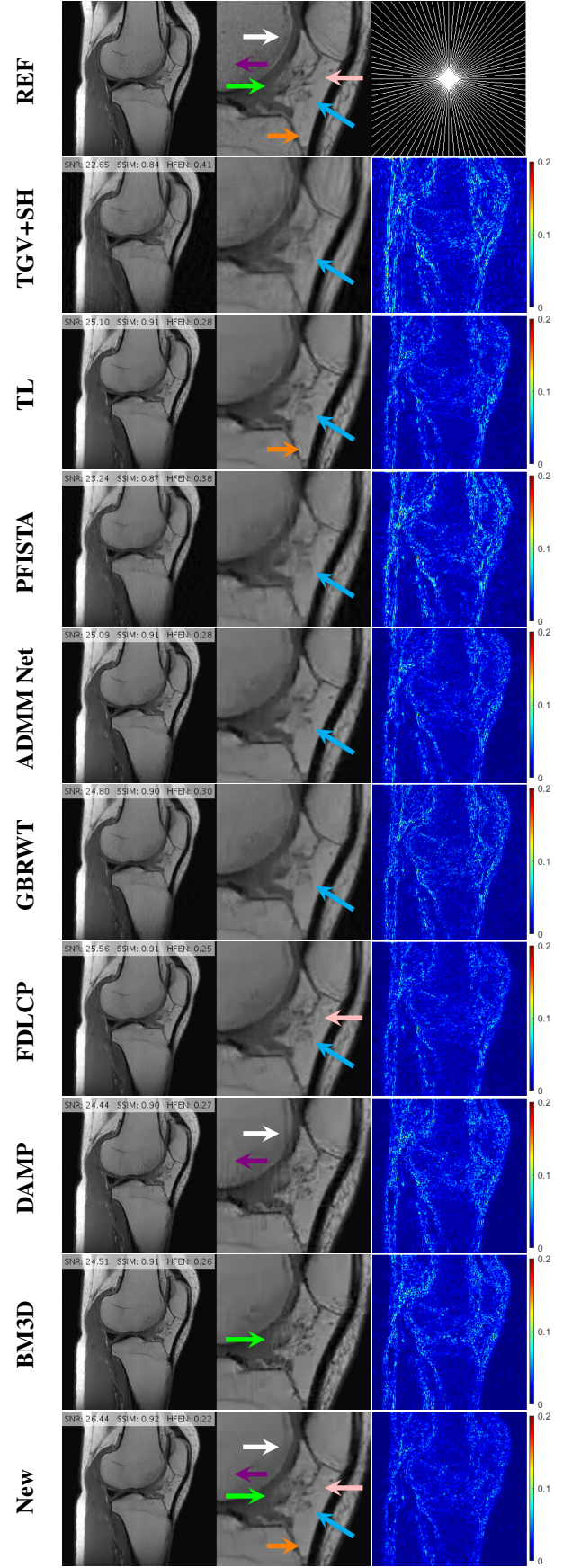

Fig. \ref{axial} demonstrates various reconstructions for an axial $T_2$-weighted brain image under a 20\% Cartesian sampling. The magnified views show that FDLCP has smoothed out fine cerebral details while DAMP, ADMM Net, GBRWT, pFISTA, TL and TGV+Sh have introduced severe artifacts into the image, not least among them is a very large dark hole in the parietal lobe.  BM3D provides a better result but this method also degrades the ventricles. The proposed framework again gives the best result with the cleanest error map.

 For convenience, in Figs. \ref{sag}, \ref{knee} and \ref{axial} we have inserted the SNR, SSIM and HFEN values for each method to the top of the corresponding reconstructed image. Moreover, some reconstruction errors in compared methods and their corrections in the proposed method are annotated with arrows.

 Some HFEN, SSIM and SNR plots related to above reconstructions are depicted in Figs. \ref{sag SNR plot}, \ref{knee SNR plot} and \ref{axial HFEN plot}. The common characteristic in these plots and many others which are not shown here for space considerations, is the fact that the BM3D-MRI algorithm stagnates and starts to compromise the solution after about 140 iterations whereas the proposed framework consistently and steadily converges to its solution. We observed that this behavior from the BM3D-MRI method is irrelevant to the number of iterations; even with 100 iterations this method would compromise the solution after about 70 iterations (see Fig. \ref{term contribution SNR plots}). We remark that the number of iterations was set to 200 in Figs. \ref{sag SNR plot}, \ref{knee SNR plot} and \ref{axial HFEN plot} (matrix size $256\times 256$) to guarantee the best performance of the other methods as well as to demonstrate the convergence and stability of the proposed framework. Currently our non-optimized, non-parallelized proof-of-principle code runs for an average of 0.65 seconds per iteration for matrix size $256\times256$.

 \begin{figure}[] \includegraphics*[height=4cm,width=0.495\linewidth]{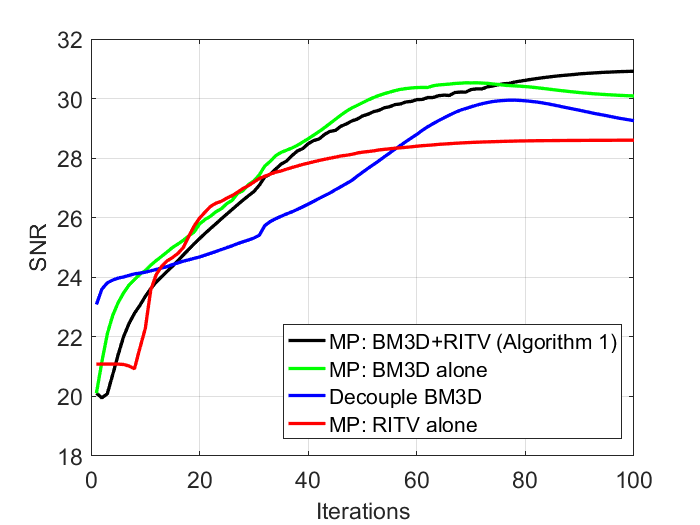}\hspace{0pt} \includegraphics*[height=4cm,width=0.495\linewidth]{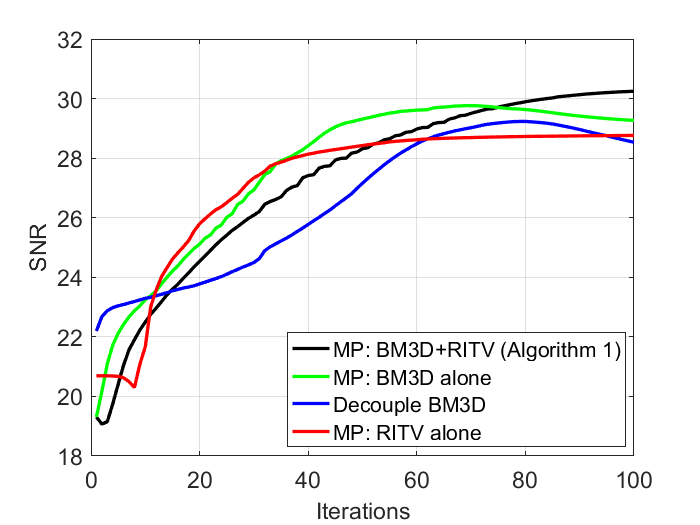}\hspace{0pt}    \caption{SNR plots for various variants of the proposed method and the decoupled BM3D-MRI \cite{BM3D-MRI} for two example slices from the test set.}\label{term contribution SNR plots}    \end{figure}

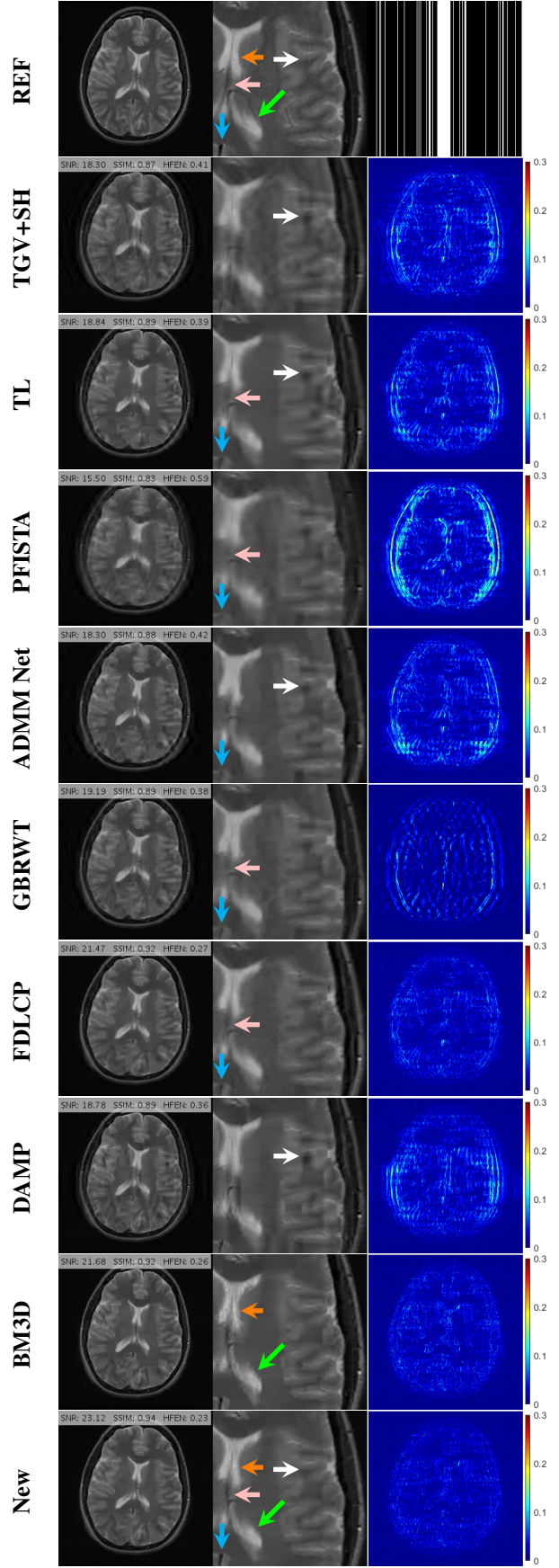
\begin{figure}\centering
 \begin{tikzpicture} \begin{scope}[] \node[anchor=north west,inner sep=0] (image) at (0,0) {  \begin{tabular}{l r} \rotatebox[origin=c]{90}{\textbf{REF}}  &  \includegraphics[scale=0.25,valign = m]{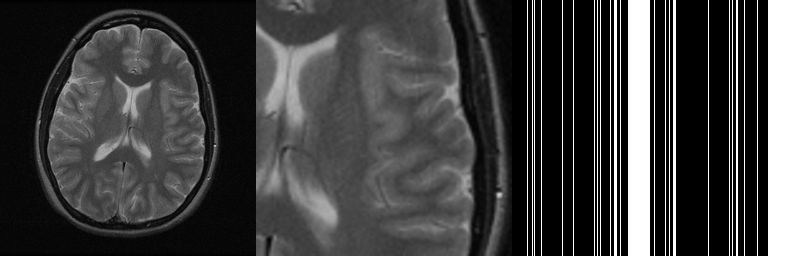}   \end{tabular}};  \node [anchor=north] (water1) at (3.27,-1.4) {}; \node [anchor=north] (water2) at (3.95,-1.1) {};\draw [-stealth, line width=2pt, pink] (water2) --++ (-.5,0);  \draw [-stealth, line width=2pt, cyan] (water1) --++ (0,-.5);
 \node [anchor=north] (water6) at (3.9,-.73) {}; \draw [-stealth, line width=2pt, white] (water6) --++ (0.5,0); \node [anchor=north] (water4) at (4,-0.7) {}; \node [anchor=north] (water5) at (4.3,-1.07) {};\draw [-stealth, line width=2pt, orange] (water4) --++ (-0.45,0); \draw [-stealth, line width=2pt, green] (water5) --++ (-0.5,-.5);\end{scope}\end{tikzpicture}%
            
\begin{tikzpicture}  \begin{scope}[] \node[anchor=north west,inner sep=0] (image) at (0,0) { \begin{tabular}{l r} \rotatebox[origin=c]{90}{\textbf{TGV+SH}}  &   \includegraphics[scale=0.25,valign = m]{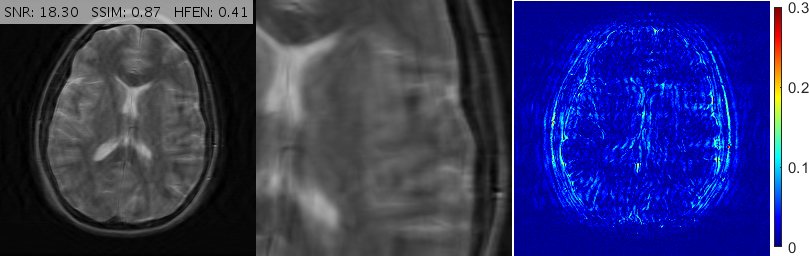} \end{tabular}};  \node [anchor=north] (water6) at (3.9,-.73) {}; \draw [-stealth, line width=2pt, white] (water6) --++ (0.5,0); 
\end{scope} \end{tikzpicture}%
            
\begin{tikzpicture}\begin{scope}[] \node[anchor=north west,inner sep=0] (image) at (0,0) { \begin{tabular}{l r} \rotatebox[origin=c]{90}{\textbf{TL}} & \includegraphics[scale=0.25,valign = m]{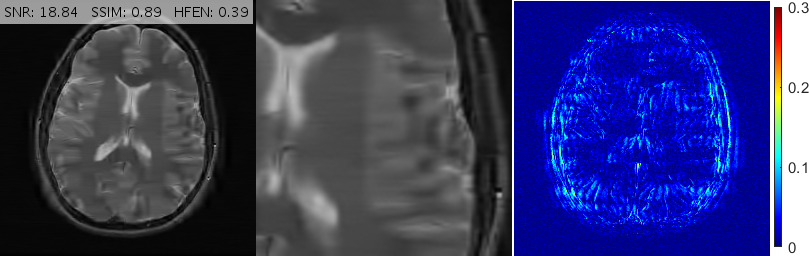} \end{tabular}};  \node [anchor=north] (water1) at (3.27,-1.4) {}; \node [anchor=north] (water2) at (3.95,-1.1) {};\draw [-stealth, line width=2pt, pink] (water2) --++ (-.5,0);  \draw [-stealth, line width=2pt, cyan] (water1) --++ (0,-.5);
\node [anchor=north] (water6) at (3.9,-.73) {}; \draw [-stealth, line width=2pt, white] (water6) --++ (0.5,0); \end{scope} \end{tikzpicture}%
            
\begin{tikzpicture}  \begin{scope}[] \node[anchor=north west,inner sep=0] (image) at (0,0) {\begin{tabular}{l r}  \rotatebox[origin=c]{90}{\textbf{PFISTA}}   &  \includegraphics[scale=0.25,valign = m]{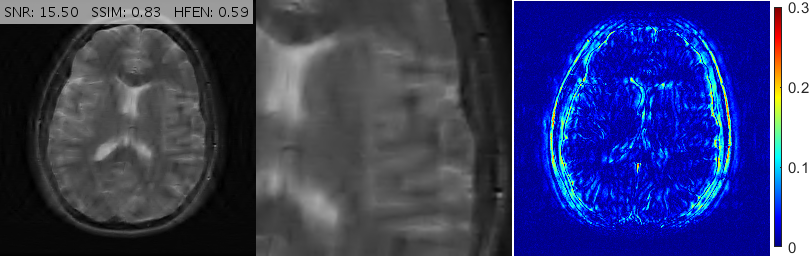} \end{tabular}};  \node [anchor=north] (water1) at (3.27,-1.4) {}; \node [anchor=north] (water2) at (3.95,-1.1) {};\draw [-stealth, line width=2pt, pink] (water2) --++ (-.5,0);  \draw [-stealth, line width=2pt, cyan] (water1) --++ (0,-.5);
\end{scope} \end{tikzpicture}%
            
\begin{tikzpicture}\begin{scope}[] \node[anchor=north west,inner sep=0] (image) at (0,0) { \begin{tabular}{l r} \rotatebox[origin=c]{90}{\textbf{ADMM Net}} & \includegraphics[scale=0.25,valign = m]{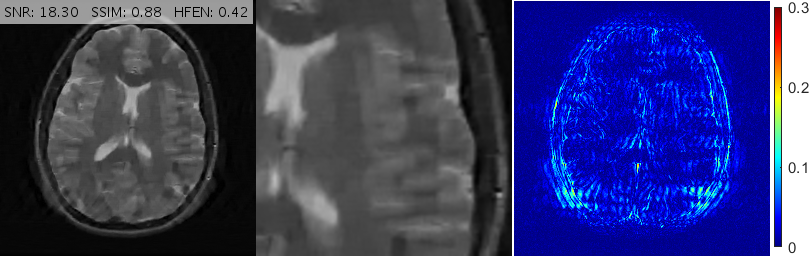} \end{tabular}}; \node [anchor=north] (water1) at (3.27,-1.4) {}; \draw [-stealth, line width=2pt, cyan] (water1) --++ (0,-.5); \node [anchor=north] (water6) at (3.9,-.73) {}; \draw [-stealth, line width=2pt, white] (water6) --++ (0.5,0); \end{scope} \end{tikzpicture}%
            
\begin{tikzpicture}\begin{scope}[] \node[anchor=north west,inner sep=0] (image) at (0,0) { \begin{tabular}{l r} \rotatebox[origin=c]{90}{\textbf{GBRWT}}  &  \includegraphics[scale=0.25,valign = m]{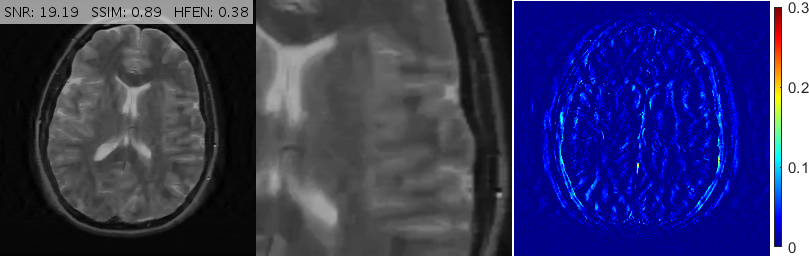} \end{tabular}};\node [anchor=north] (water1) at (3.27,-1.4) {}; \node [anchor=north] (water2) at (3.95,-1.1) {};\draw [-stealth, line width=2pt, pink] (water2) --++ (-.5,0);  \draw [-stealth, line width=2pt, cyan] (water1) --++ (0,-.5); \end{scope} \end{tikzpicture}%
            
\begin{tikzpicture} \node [anchor=north] (water1) at (3.27,-1.4) {}; \node [anchor=north] (water2) at (3.95,-1.1) {}; \begin{scope}[] \node[anchor=north west,inner sep=0] (image) at (0,0) {\begin{tabular}{l r} \rotatebox[origin=c]{90}{\textbf{FDLCP}}  &    \includegraphics[scale=0.25,valign = m]{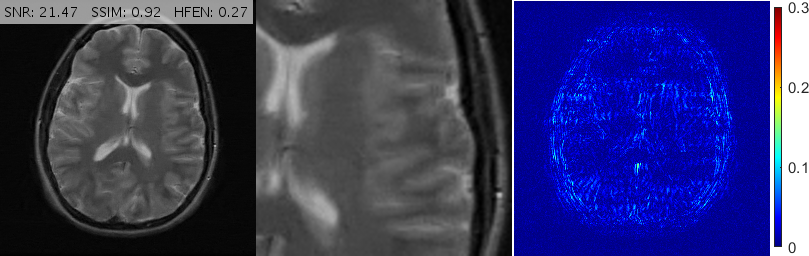}  \end{tabular}}; \draw [-stealth, line width=2pt, pink] (water2) --++ (-.5,0);  \draw [-stealth, line width=2pt, cyan] (water1) --++ (0,-.5);  \end{scope} \end{tikzpicture}%
            
 \begin{tikzpicture}  \begin{scope}[] \node[anchor=north west,inner sep=0] (image) at (0,0) {\begin{tabular}{l r} \rotatebox[origin=c]{90}{\textbf{DAMP}}  &    \includegraphics[scale=0.25,valign = m]{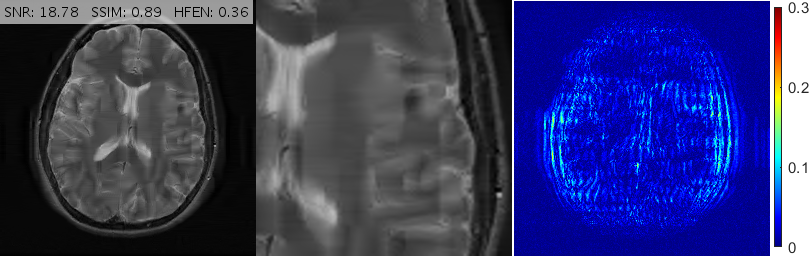}  \end{tabular}}; \node [anchor=north] (water6) at (3.9,-.73) {}; \draw [-stealth, line width=2pt, white] (water6) --++ (0.5,0);    \end{scope} \end{tikzpicture}%
            
\begin{tikzpicture}  \begin{scope}[] \node[anchor=north west,inner sep=0] (image) at (0,0) { \begin{tabular}{l r} \rotatebox[origin=c]{90}{\textbf{BM3D}}  &  \includegraphics[scale=0.25,valign = m]{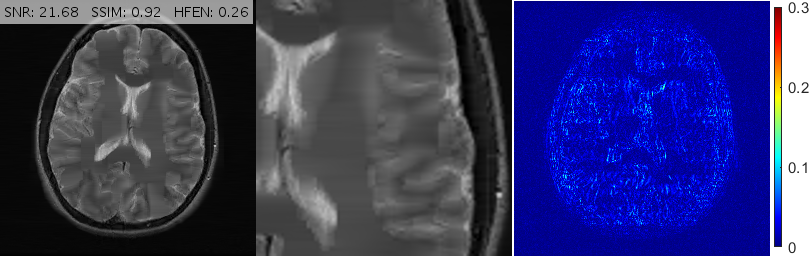}  \end{tabular}};  \node [anchor=north] (water4) at (4,-0.7) {}; \node [anchor=north] (water5) at (4.3,-1.07) {};\draw [-stealth, line width=2pt, orange] (water4) --++ (-0.45,0); \draw [-stealth, line width=2pt, green] (water5) --++ (-0.5,-.5);  \end{scope} \end{tikzpicture}%
            
\begin{tikzpicture}  \begin{scope}[] \node[anchor=north west,inner sep=0] (image) at (0,0) {\begin{tabular}{l r} \rotatebox[origin=c]{90}{\textbf{New}}  &    \includegraphics[scale=0.25,valign = m]{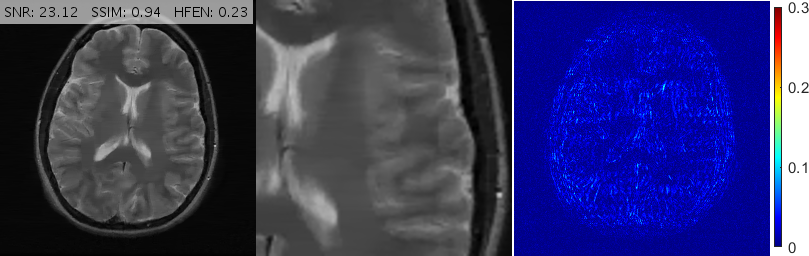}  \end{tabular}}; \node [anchor=north] (water1) at (3.27,-1.4) {}; \node [anchor=north] (water2) at (3.95,-1.1) {};\draw [-stealth, line width=2pt, pink] (water2) --++ (-.5,0);  \draw [-stealth, line width=2pt, cyan] (water1) --++ (0,-.5);
\node [anchor=north] (water6) at (3.9,-.73) {}; \draw [-stealth, line width=2pt, white] (water6) --++ (0.5,0); \node [anchor=north] (water4) at (4,-0.7) {}; \node [anchor=north] (water5) at (4.3,-1.07) {};\draw [-stealth, line width=2pt, orange] (water4) --++ (-0.45,0); \draw [-stealth, line width=2pt, green] (water5) --++ (-0.5,-.5); \end{scope} \end{tikzpicture}%
             
\caption{Comparison between various reconstructions for a $256\times256$ $T_2$-weighted TSE axial brain image with 20\% Cartesian sampling.  The left, middle and right columns respectively show the reconstructions, magnified views and error maps. }\label{axial}
\end{figure}

 {  \subsection{Contribution of each component of the proposed method in overall performance} \label{sec term contribution}
 }

   The proposed framework involves three ingredients: The RITV term, the BM3D term and the Malitsky-Pock (MP) algorithm where the two terms meet. Here we dissect the contribution of each ingredient in the overall performance of the proposed method. In order to do this, we turn off the BM3D term and the RITV term (with the constraint) in turn in (\ref{proposed model}) and make the necessary adjustments to Algorithm \ref{proposed algorithm}. We (randomly) select 20 slices from our knee data with the 16\% spiral sampling pattern from  Subsection \ref{other works} to perform the analysis. All the parameters are borrowed from Subsection \ref{sec setup}, with the only exception being made when the BM3D term is turned off, in which case we set $\beta = 10^{-5}$ to speed up the convergence of the RITV-only variant.

   The SNR plots for two example slices are given in Fig. \ref{term contribution SNR plots}. The observations can be summarized as follows:
   \begin{enumerate}
   \item When RITV is removed (turned off) in (\ref{proposed model}), the resulting MP: BM3D-only algorithm practically solves the same model as BM3D-MRI \cite{BM3D-MRI}. The Malitsky-Pock algorithm outperforms the decoupled algorithm of \cite{BM3D-MRI}, however, a similar stagnating effect as that of the Decoupled BM3D-MRI is observed in MP: BM3D-only, albeit to a lesser severity, starting near the 70th iteration. 
   \item When the BM3D term is removed from (\ref{proposed model}), the resulting MP: RITV-only algorithm is on average outperformed by the BM3D-based methods, however, it is remarkably stable with no stagnation visible within 100 iterations.
   \item When both BM3D and RITV terms are turned on (which simply results in Algorithm (\ref{proposed algorithm}), alternatively labeled MP: BM3D+RITV), not only is the stagnation of MP: BM3D-only  and Decoupled BM3D-MRI resolved, but the overall SNR is also improved.
   \end{enumerate} 
The observations above are numerically reported in Table \ref{Table3}. The SNR drop should be viewed as a metric of stagnation; the larger the SNR drop, the more severe the stagnation.

\begin{table}\centering
         \caption{SNR behavior for variants of the proposed methed and the Decoupled BM3D-MRI method.  Maximum SNR is the highest SNR among all iteration. Final SNR is the SNR at the last (100th) iteration. SNR drop is the difference between Max SNR and Final SNR. The entries of the table have been averaged over 20 experiments. }
        \begin{tabular}{lccc}
        \toprule Method & Max SNR  & Final SNR & SNR drop \\ \hline
        MP: BM3D+RITV & 29.40 & 29.40 & 0.00 \\
        MP: BM3D alone & 29.13 & 28.62 & 0.51 \\ Decoupled BM3D & 28.62 & 
        27.84 & 0.78 \\ MP: RITV alone & 27.22 & 27.22  & 0.00 
        \\ \bottomrule
        \label{Table3}
        \end{tabular}
         \end{table}

{\subsection{Comparison with other works: noisy scenario}\label{noisy scenario} }

\begin{figure}\centering
\includegraphics*[scale=0.25]{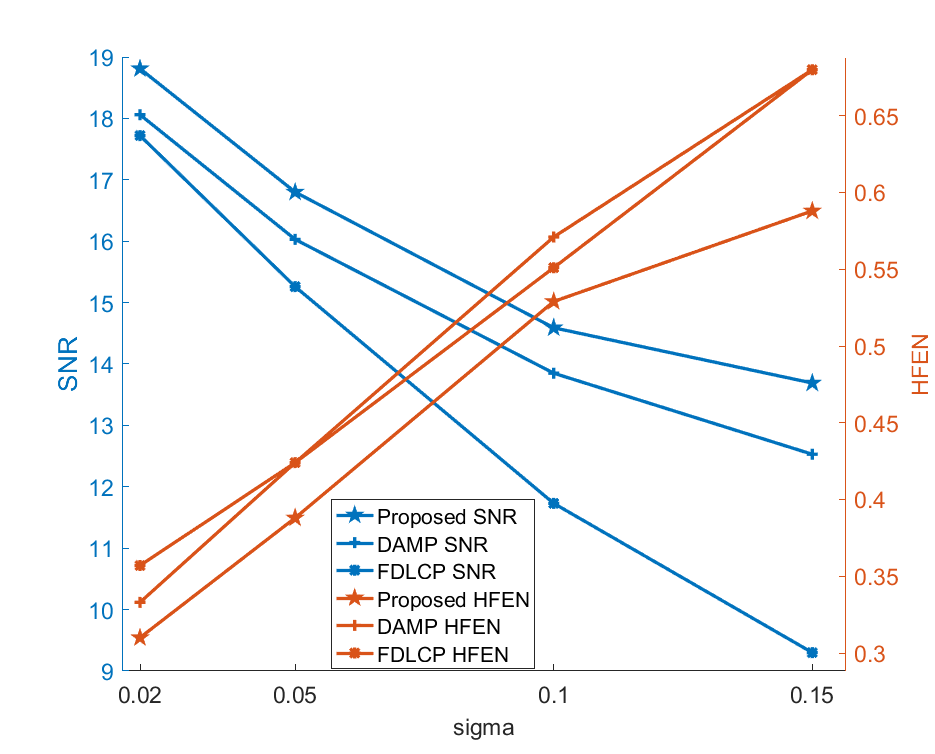}
\caption{Performance of the proposed compared with DAMP and FDLCP in noisy scenario. Each point on the graph is averaged over 50 experiments.}
\label{noisy_experiments}
\end{figure}

In Subsection \ref{other works} the performance of the proposed method was compared with other state-of-the-art methods in the noise-free case. In this subsection we investigate how our method performs in noisy settings.

Since most of the compared works in Subsection \ref{other works} only consider noiseless scenario, and to make the current subsection concise and to the point, we only provide comparisons with \textbf{DAMP} \cite{DAMP} which is an extension of BM3D-MRI that considers noisy scenario as well, and \textbf{FDLCP} \cite{FDLCP} which demonstrated a consistently decent performance in noiseless experiments.

We select our brain data of 50 slices for the experiments and again, resize them to $128\times128$ for speed. The Fourier spectrum of each slice is corrupted by additive white Gaussian noise with standard deviations $\sigma = 0.02, 0.05, 0.1$ and $0.15$. Four random noise masks corresponding to above $\sigma$'s are generated. Once created, we fix the masks to make fair comparisons between the three methods. This approach was also used in \cite{DAMP}. We also fix a Cartesian pattern of 30\% sampling for the experiments.

In accordance with noise levels above, in the proposed method we set $\eta = 8,\,14,\,30$ and $38$, and $\beta=0.1,\,1$ for the first two levels and $\beta=7$ for the last two.  As a general rule, $\eta$ and $\beta$ increase with noise level (however, for matrix size $256\times256$ simply fixing $\beta=0.1$ for all $\sigma$'s is sufficient).

In the DAMP paper \cite{DAMP} the parameter $\lambda_{\text{3D}}$ is suggested to remain fixed at $2.7$ for noisy experiments. However, We observed empirically that this algorithm performs better if we set $\lambda_{\text{3D}}$ according to noise levels as $2.6, 2.7, 3$ and $3.5$. We make these choices instead of the fixed value above.

The FDLCP paper \cite{FDLCP} does not consider a noisy scenario. Nevertheless, we observed that for the noise levels above, this method performs better with parameters accordingly set as $\beta = 2^4, 2^3, 2^2, 2^1$ and $\lambda = 10^5$ for the first three noise levels and $\lambda = 10^6$ for $\sigma=0.15$.
   
 
 \begin{figure}\centering

\begin{tikzpicture} \begin{scope}[] \node[anchor=north west,inner sep=0] (image) at (0,0) { \begin{tabular}{l r} \rotatebox[origin=c]{90}{\textbf{REF \& ZF}}  &   \includegraphics[scale=0.28 ,valign = m]{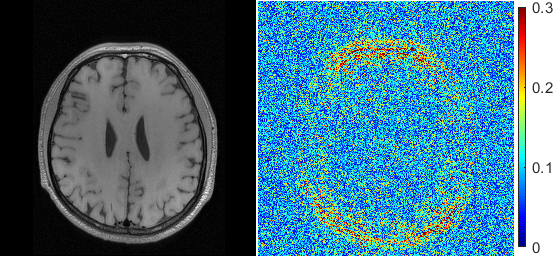} \end{tabular}}; \end{scope} \end{tikzpicture}%

\begin{tikzpicture} \begin{scope}[] \node[anchor=north west,inner sep=0] (image) at (0,0) { \begin{tabular}{l r} \rotatebox[origin=c]{90}{\textbf{FDLCP}}  &   \includegraphics[scale=0.28 ,valign = m]{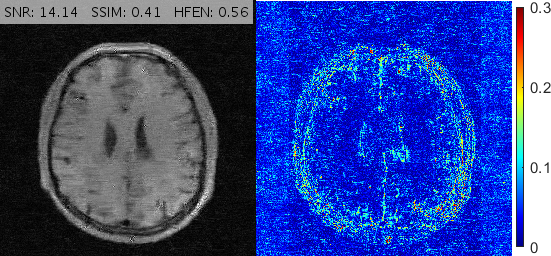} \end{tabular}}; \end{scope} \end{tikzpicture}%

\begin{tikzpicture} \begin{scope}[] \node[anchor=north west,inner sep=0] (image) at (0,0) { \begin{tabular}{l r} \rotatebox[origin=c]{90}{\textbf{DAMP}}  &   \includegraphics[scale=0.28 ,valign = m]{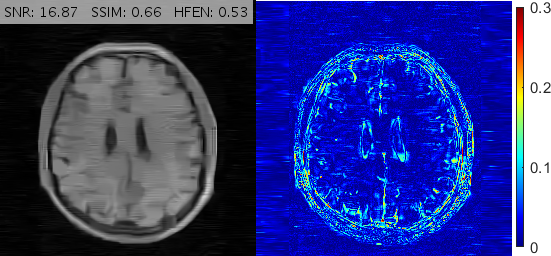} \end{tabular}}; \end{scope} \end{tikzpicture}%

\begin{tikzpicture} \begin{scope}[] \node[anchor=north west,inner sep=0] (image) at (0,0) { \begin{tabular}{l r} \rotatebox[origin=c]{90}{\textbf{New}}  &   \includegraphics[scale=0.28 ,valign = m]{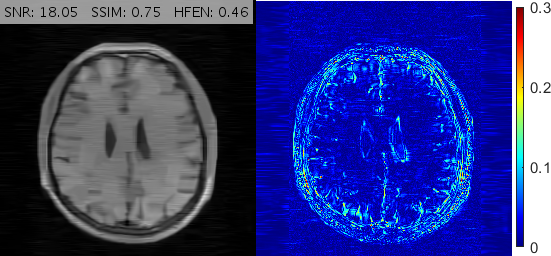} \end{tabular}}; \end{scope} \end{tikzpicture}%
\caption{Comparison between various reconstructions for an example $T_1$-weighted $256\times256$ slice with Cartesian sampling at \%20 and noise standard deviation $\sigma = 0.1$. The top row shows the reference image and the error map of the initial noisy zero-filling solution.} \label{noisy}
\end{figure}
 The overall performance of the three methods in all 50 brain slices is presented in Fig. \ref{noisy_experiments}. Furthermore, to obtain a visual idea of the performances, Fig. \ref{noisy} shows an example $256\times256$-sized $T_1$-weighted slice reconstructed from 20\% sampling with $\sigma = 0.1$.



\section{Conclusions}\label{sec concl}

We observed that subjecting the total variation functional to constraints that improve its rotation-invariance property can be significantly useful in compressed MRI reconstruction. One might expect that a similar modification for TGV can potentially improve TGV-based solutions as well. The proposed framework may also be extended to multi-modality imaging techniques such as MR-PET. Work in this direction is underway.  
\section*{Acknowledgment}
This research did not receive any specific grant from funding agencies in the public, commercial, or not-for-profit sectors. The graciousness of all researchers and authors of cited works is appreciated for sharing their data and software. Perceptive comments from anonymous reviewers are also appreciated. During the final preparations of this paper, a number of very young and very talented students from our community lost their lives in a tragic plane crash. This work is dedicated to the living memory of our friends and all the victims of that tragedy.

{
\appendix \section{Proof of Theorem \ref{theorem}}  \label{proof of theorem}
}
 Assume $u, v_s, s\in S$ is a solution of problem (\ref{proposed model}) with $\eta=0$. Consider the problem 
   \begin{equation} \label{2proposed model}
   \begin{split}
   \min_{\bar{u},\{\bar{v}_s:s\in S\}}
   \frac{1}{2}||&\mathcal{F}_\mathcal{M}\bar{u}-\mathcal{F}(\mathcal{R}u_{zf})||_2^2 + 
   \lambda\sum_{s\in S} ||\bar{v}_s||_{1,2}\\
   &\text{s.t. } \sum_{s\in S} L^*_s\bar{v}_s - D\bar{u} = 0,
   \end{split}
   \end{equation}
   we will show that a solution of problem (\ref{2proposed model}) is as follows:
   $$\bar{u}=\mathcal{R}u, \bar{v}_\bullet=(\bar{v}^1_\bullet, \bar{v}^2_\bullet)=(-\mathcal{R}{v}^2_\bullet, \mathcal{R}{v}^1_\bullet)$$ $$\bar{v}_\updownarrow(i,j)=(\bar{v}^1_\updownarrow(i,j), \bar{v}^2_\updownarrow(i,j))=(-\mathcal{R}{v}^2_\leftrightarrow(i+1,j), \mathcal{R}{v}^1_\leftrightarrow(i+1,j)),
   $$
   $$\bar{v}_\leftrightarrow=(\bar{v}^1_\leftrightarrow, \bar{v}^2_\leftrightarrow)=(-\mathcal{R}{v}^2_\updownarrow, \mathcal{R}{v}^1_\updownarrow), \bar{v}_+=(\bar{v}^1_+, \bar{v}^2_+)=(-\mathcal{R}{v}^2_+, \mathcal{R}{v}^1_+).$$
   For this reason, it is enough to show that the above solution satisfies the constraints and the optimal objective value is equal to the optimal objective value of problem (\ref{proposed model}) with sampled k-space $b$. Note that a smaller objective value is not possible because if it were, then with a similar argument we could get a better solution for the original problem which is a contradiction.
   
  {From the definition of adjoint operators and the fact that for any two dimensional discrete complex image $A\in\mathbb{C}^{n\times n},\, \mathcal{R}A(j,i)=A(i,n-j+1),$
   we get
   \begin{equation}\label{1}
   \sum_{s\in S} L^*_s\bar{v}_s - D\bar{u} = 0,
   \end{equation} 
   or equivalently 
       \begin{equation}\label{1.1}
       	\sum_{s\in S} (L^*_s\bar{v}_s)_k - (D\bar{u})_k = 0, k=1,2.
       \end{equation}
       We will show (\ref{1.1}) for $k=1$. For $k=2$ the argument is similar.  \\
       From the definition of adjoint operator, it is observed that       
\begin{equation*}
\begin{split}
(L_{\bullet}^*v_{\bullet})_1(i,j)&=\frac{1}{2}\left\{v_{\bullet}^1(i,j)+v_{\bullet}^1(i+1,j)\right\},\\
(L_{\bullet}^*v_{\bullet})_2(i,j)&=\frac{1}{2}\left\{v_{\bullet}^2(i,j)+v_{\bullet}^2(i,j+1)\right\},\\
(L_{\updownarrow}^*v_{\updownarrow})_1(i,j)&=v_{\updownarrow}^1(i,j),\\
(L_{\updownarrow}^*v_{\updownarrow})_2(i,j)&=\frac{1}{4}\big(v_{\updownarrow}^2(i,j)+v_{\updownarrow}^2(i,j+1)+v_{\updownarrow}^2(i-1,j)\\&+v_{\updownarrow}^2(i-1,j+1) \big),\\
(L_{\leftrightarrow}^*v_{\leftrightarrow})_1(i,j)&= \frac{1}{4}\big(v_{\leftrightarrow}^1(i,j)+v_{\leftrightarrow}^1(i+1,j)+v_{\leftrightarrow}^1(i,j-1)\\&+v_{\leftrightarrow}^1(i+1,j-1)\big),\\
(L_{\leftrightarrow}^*v_{\leftrightarrow})_2(i,j)&=v_{\leftrightarrow}^2(i,j),\\  
(L_{+}^*v_{+})_1(i,j)&=\frac{1}{2}(v_+^1(i,j)+v_+^1(i,j-1)),\\  
(L_{+}^*v_{+})_2(i,j)&=\frac{1}{2}(v_+^2(i,j)+v_+^2(i-1,j)).  
\end{split}
\end{equation*}
From the definition of difference operator we get
       $$(D\bar{u})_1(j,i)=u(i,n-j)-u(i,n-j+1),$$
       On the other hand
       $$\bar{v}^1_s(j,i)=-v^2_s(i,n-j+1);\, s\in \{\bullet,+\},$$
   $$\bar{v}^1_{\updownarrow}(j,i)=-v^2_\leftrightarrow (i,n-j),\,\bar{v}^1_{\leftrightarrow}(j,i)=-v^2_\updownarrow (i,n-j+1).$$
       Consequently
   \begin{equation*}
    \begin{split}
    (L_{\bullet}^*\bar{v}_{\bullet})_1(j,i)&=-\frac{1}{2}\left\{v_{\bullet}^2(i,n-j+1)+v_{\bullet}^2(i,n-j)\right\}\\&=-(L_{\bullet}^*{v}_{\bullet})_2(i,n-j).
       \end{split}
      \end{equation*}
             Similarly we have the following relations:
             $$(L_{\updownarrow}^*\bar{v}_{\updownarrow})_1(j,i)=-(L_{\leftrightarrow}^*{v}_{\leftrightarrow})_2(i,n-j),$$
                         \begin{equation}\label{prf}
                         (L_{\leftrightarrow}^*\bar{v}_{\leftrightarrow})_1(j,i)=-(L_{\updownarrow}^*{v}_{\updownarrow})_2(i,n-j),
                         \end{equation}
                 $$(L_{+}^*\bar{v}_{+})_1(j,i)=-(L_{+}^*{v}_{+})_2(i,n-j),$$
                 $$(D\bar{u})_1(j,i)=-(D{u})_2(i,n-j).$$          
                From the original problem (\ref{proposed model}), we know that the following constraint holds
                 $$\sum_{s\in S} L^*_sv_s - Du = 0.$$
                 Therefore from equalities in (\ref{prf}), equation (\ref{1.1}) is proved for $k=1.$
   Moreover, using the $l_{1,2}$ norm definition (\ref{lp norm}) along with zero boundary conditions of the gradient field variables (Subsection \ref{RITV intro}), it is easily seen that  
  $\sum_{s\in S} ||\bar{v}_s||_{1,2} = \sum_{s\in S} ||v_s||_{1,2}$, hence $\text{RITV}(\mathcal{R}{u}) = \text{RITV}(u)$.
    It remains to show that
   \begin{equation}\label{3}
   ||\mathcal{F}_\mathcal{M}\mathcal{R}{u}-\mathcal{F}(\mathcal{R}u_{zf})||_2^2=||\mathcal{F}_\mathcal{M}{u}-\mathcal{F}u_{zf}||_2^2,
   \end{equation}
   equivalently 
   $$||\mathcal{M}\odot\mathcal{F} (\mathcal{R}{u})-\mathcal{F}(\mathcal{R}u_{zf})||_2^2=||\mathcal{F}_\mathcal{M}{u}-\mathcal{F}u_{zf}||_2^2.$$
   As k-space data $\mathcal{F}(u_{zf})$ and $\mathcal{F}(\mathcal{R}u_{zf})$ are obtained by mask 
   $\mathcal{M},$ the equal form of (\ref{3}) is
   $$||\mathcal{M}\odot\mathcal{F} (\mathcal{R}({u-u_{zf}}))||_2^2=||\mathcal{M}\odot\mathcal{F}({u}-u_{zf})||_2^2.$$
   By the norm-preserving assumption on $\mathcal{M},$ {it is enough to show that 
   $$\|\mathcal{F}(\mathcal{R}w)\|_2=\|\mathcal{F}w\|_2, w\in\mathbb{C}^{n\times n}.$$}
   It is well known that $\mathcal{F}w=(F_n w F_n)$ where 
   $F_n$ is the Fourier matrix, given by $(F_n)(i,j)=\frac{1}{n}(\omega)^{ij}$ with $\omega=e^{-\frac{2\pi i}{n}}$. Thus
   \begin{equation}
   (\mathcal{F}w)_{i,j}=\sum_s\sum_k (F_n)_{i,k} w_{k,s} (F_n)_{s,j}.
   \end{equation}
   Now, we obtain $\mathcal{F}(\mathcal{R}w)$ and $\mathcal{R}(\mathcal{F}w);$
   $$\mathcal{F}(\mathcal{R}w)_{j,i}=\sum_s\sum_k (F_n)_{j,s} w_{k,n-s+1} (F_n)_{k,i}=$$
   \begin{equation}\label{4}
   \sum_s\sum_k (F_n)_{i,k} w_{k,s} (F_n)_{n-s+1,j}=\sum_s\sum_k (F_n)_{i,k} w_{k,s} (F_n)_{-s+1,j}.
   \end{equation}
   Moreover, 
   $$\mathcal{R}(\mathcal{F}w)(j,i)=\sum_s\sum_k (F_n)_{i,k} w_{k,s} (F_n)_{s,n-j+1}=$$
   \begin{equation}\label{5}
   \sum_s\sum_k (F_n)_{i,k} w_{k,s} (F_n)_{s,-j+1}.
   \end{equation}
   From (\ref{4}) and (\ref{5}) we get
   $$\mathcal{R}(\mathcal{F}w)_{j+1,i}=\omega^j \mathcal{F}(\mathcal{R}w)_{j,i};\, i,j\in\{0,\cdots,n-1\}.$$
   Since $|\omega^j|=1$ and $\mathcal{R}(\mathcal{F}w)_{n,i}=\mathcal{R}(\mathcal{F}w)_{0,i}$ we have 
   $$\|\mathcal{F}w\|_2=\|\mathcal{R}(\mathcal{F}w)\|_2=\|\mathcal{F}(\mathcal{R}w)\|_2,$$
   and the proof is complete. $\quad \square$\\
   \\


\begin{thebibliography}{99} 
 
   \bibitem{CS} D. Donoho,                       {\em Compressed sensing,}    IEEE Transactions on Information Theory, vol. 52, no. 4,
             pp. 1289-1306, 2006.
             
 \bibitem{LDP} M. Lustig, D. Donoho and J. pauly,            {\em Sparse MRI: The application of compressed
 sensing for rapid MR imaging,} Magnetic Resonance in Medicine, vol. 58, no. 6, pp. 1182-1195, 2007.
 
  \bibitem{Ma MRI} S. Ma, W. Yin, Y. Zhang and A. Chakraborty,                {\em An efficient algorithm for
           compressed MR imaging using total variation and wavelets,} in Proceedings of CVPR, pp. 1-8, 2008.
             
   \bibitem{FCSA} J. Huang, S. Zhang and D. Metaxas, {\em Efficient MR image reconstruction for
       compressed MR imaging,}    Medical Image Analysis, vol. 15, no. 5, pp. 670-679, 2011.
    
 \bibitem{CS MRI} M. Lustig, D. Donoho, J. Santos, and J. pauly,  {\em Compressed
 Sensing MRI,}  IEEE Signal Processing Magazine, vol. 25, no. 2, pp. 72-82, 2008.
 \bibitem{TGV} K. Bredies, K. Kunisch and T. Pock,        {\em Total generalized variation,}    SIAM Journal on Imaging Sciences, vol. 3, no. 3, pp. 492-526, 2010.
  \bibitem{TGV MRI} F. Knoll, K. Bredies, T. Pock and R. Stollberger,               {\em Second order total generalized variation (TGV) for MRI,}  Magnetic Resonance in Medicine, vol. 65, no. 2, pp. 480-491, 2011.
 \bibitem{TGVSH} W. Guo, J.Kin and W.Yin, {\em A new detail-preserving regularization scheme,} SIAM Journal on Imaging Sciences, vol. 7, no. 2, pp. 1309-1334, 2014.
          
 \bibitem{DLMRI}
   S. Ravishankar, Y. Bresler,
          {\em MR image reconstruction from highly undersampled k-space data by
          dictionary learning,}
          IEEE Transactions on Medical Imaging, vol. 30, no. 5, pp. 1029-1041, 2010.
          
          \bibitem{TLMRI}
                    S. Ravishankar, Y. Bresler,
                    {\em Efficient blind compressed sensing using
                    sparsifying   transforms  with  convergence   guarantees   and  application
                    to  magnetic  resonance  imaging,}
                   SIAM  Journal  on  Imaging  Sciences, vol. 8, no. 4, pp. 2519-2557, 2015.
                    
  \bibitem{Beck} A. Beck,           {\em First-Order Methods in Optimization,}    SIAM, Philadelphia, 2017.
       
 \bibitem{Cham-Pock} A. Chambolle and T. Pock,                            {\em A first-order primal-dual algorithm for
        convex problems with applications to imaging,} Journal of Mathematical
        Imaging and Vision, vol. 40, no. 1, pp. 120-145, 2010.
        
\bibitem{FRIST}
 B. Wen, S. Ravishankar and Y. Bresler,
 {\em FRIST-flipping and rotation invariant sparsifying transform learning and applications,}
 Inverse Problems, vol. 33, no. 7, pp. 074007, 2017.
        
 \bibitem{BM3D 1}
  K. Dabov, A. Foi, V. Katkovnik and K. Egiazarian,
  {\em Image denoising by sparse 3D transform-domain
  collaborative filtering,}
  IEEE Transactions on Image Processing, vol.16, no.8, pp. 2080-2095, 2007.               
           
 \bibitem{BM3D 2}
  A. Danielyan, V. Katkovnik and K. Egiazarian,
  {\em BM3D Frames and Variational Image Deblurring,}
  IEEE Transactions on Image Processing, vol. 21, no. 4, pp. 1715-1728, 2012.               
                    
 \bibitem{BM3D-MRI}
  E. M. Eksioglu,
  {\em Decoupled algorithm for MRI reconstruction using nonlocal block matching model: BM3D-MRI,}
  Journal of Mathematical Imaging and Vision, vol. 56, no. 3, pp. 430-440, 2016.               
 
 \bibitem{MP}
   Y. Malitsky and T. Pock,
   {\em A first-order primal-dual algorithm with linesearch,}
   SIAM Journal on Optimization, vol. 28, no. 1, pp. 411-432, 2018.   
 
 \bibitem{Condat}
    L. Condat,
    {\em Discrete total variation: new definition and minimization,}
    SIAM Journal on Imaging Sciences, vol. 10, no. 3, pp. 1258-1290, 2017.
    
\bibitem{ASGARD} Q. Tran-Dinh, O. Fercoq and V. Cevher, {\em  A smooth primal-dual optimization framework for nonsmooth composite convex minimization,}  SIAM Journal on Optimization, vol. 28, no. 1, pp. 96-134, 2018.     
            
    
\bibitem{pFISTA}
   Y. Liu, Z. Zhan, J. F. Cai, D. Guo, Z. Chen and X. Qu,
   {\em Projected iterative soft-thresholding algorithm for tight frames in compressed sensing magnetic resonance imaging,}
    IEEE Transactions on Medical Imaging, vol. 35, no. 9, pp. 2130-2140, 2016. 
 
\bibitem{FDLCP}
   Z. Zhan, J.-F. Cai, D. Guo, Y. Liu, Z. Chen and X. Qu,
   {\em Fast multiclass dictionaries learning with geometrical directions in MRI reconstruction,}
    IEEE Transactions on Biomedical Engineering, vol.  63, no. 9, pp. 1850-1861, 2016.  
                 
\bibitem{GBRWT}
   Z. Lai, X. Qu, Y. Liu, D. Guo, J. Ye, Z. Zhan et al, 
   {\em Image reconstruction of compressed sensing MRI using graph-based redundant wavelet transform,}
   Medical Image Analysis, vol. 27, pp. 93-104, 2016.       
   
\bibitem{Deep ADMM}
   Y. Yang, J. Sun, H. Li, and Z. Xu,
   {\em Deep ADMM-Net for compressive
   sensing MRI,}
   in Proceedings of NIPS, pp. 10-18, 2016.       
\bibitem{VN}
   K. Hammernik,
   T. Klatzer,
   E. Kobler,
   M. P. Recht,
   D. K. Sodickson,
   T. Pock
    et al, 
   {\em Learning a variational network
   for reconstruction of accelerated MRI data,}
   Magnetic Resonance in Medicine, vol. 27, no.6, pp. 3055-3071, 2018.              
\bibitem{PANO}
   X. Qu, Y. Hou, F. Lam, D. Guo, J. Zhong and Z. Chen, 
   {\em Magnetic resonance image reconstruction from undersampled measurements using a patch-based nonlocal operator,}
   Medical Image Analysis, vol. 18, no. 6, pp. 843-856, 2014.                
\bibitem{PBDW}
   X. Qu, D. Guo, B. Ning, Y. Hou, Y. Lin, S. Cai et al, 
   {\em Undersampled MRI reconstruction with patch-based directional wavelets,}
   Magnetic Resonance Imaging,  vol. 30, no. 7, pp. 964-977, 2012.      
    
      \bibitem{NYU} 
      F. Knoll et al,
  {\em  NYU machine learning data,} \url{http://mridata.org/list?project=NYU%20machine%20learning%20data}, Online, accessed September 2019. 

 \bibitem{T2} 
      E. Bullitt,
  {\em  T2,} \url{http://hdl.handle.net/1926/1148}, Online, accessed January 2020. 


 \bibitem{T1} 
      E. Bullitt,
  {\em  T1-MPRage,} \url{http://hdl.handle.net/1926/920}, Online, accessed January 2020. 

 \bibitem{DTI} 
      E. Bullitt,
  {\em  DTI,} \url{http://hdl.handle.net/1926/1178}, Online, accessed January 2020. 



   
\bibitem{DAGAN}
   G. Yang, S. Yu, H. Dong, G. Slabaugh, P. L. Dragotti, X. Ye et al, 
   {\em DAGAN: deep de-aliasing generative adversarial networks for fast compressed sensing MRI reconstruction,}
   IEEE Transactions on Medical Imaging,  vol. 37, no. 6, pp. 1310-1321, 2018.   
   
 \bibitem{GADMM}
    W. Deng and W. Yin, 
    {\em On the global and linear convergence of the generalized alternating direction method of multipliers,}
    Journal of Scientific Computing,  vol. 66, no. 3, pp. 899-916, 2016.                 
  
 
 \bibitem{Morteza}
    M. Mardani, Q. Sun, D. Donoho, V. Papyan, H. Monajemi, S. Vasanawala et al,
    {\em Neural proximal gradient descent for compressive imaging,}
    in Proceedings of NIPS, pp. 9573-9583, 2018.                  
\bibitem{DAMP}
   E. M. Eksioglu and A. K. Tanc, 
   {\em Denoising AMP for MRI reconstruction: BM3D-AMP-MRI,}
   SIAM Journal on Imaging Sciences,  vol. 11, no. 3, pp. 2090-2109, 2018.
\bibitem{Dataset}
E. Ebrahim Esfahani, {\em Data for: Compressed MRI reconstruction exploiting a rotation-invariant total variation discretization,} Mendeley Data, v1, 
\url{http://dx.doi.org/10.17632/ghfkv5w8n2.1}, 2020.         
          
                 
 \end{thebibliography}
\end{document}